\magnification=1200 
\baselineskip=19pt 
\def\Chandra{Chandrasekhar\ }
\def\Ho{H\"oflich\ }
\def\h{$H_0$ \ }
\def\h={$H_0=}

\def\ho{H\"oflich \ }

\def\mni{$M_{Ni}$ \ }
\def\mni={$M_{Ni}=}

\def\msun{$M_\odot$}

\bigskip
\bigskip
\bigskip
\centerline {TYPE Ia SUPERNOVAE AND THE HUBBLE CONSTANT}
\bigskip
\bigskip
\noindent {\sl David Branch} (branch@mail.nhn.ou.edu)
\bigskip
\noindent Department of Physics and Astronomy, University of Oklahoma,
Norman, OK 73019
\bigskip
\bigskip
\centerline {ABSTRACT}
\medskip

The focus of this review is the work that has been done during the
1990s on using Type Ia supernovae (SNe~Ia) to measure the Hubble
constant ($H_0$).  SNe~Ia are well suited for measuring $H_0$. A
straightforward maximum--light color criterion can weed out the
minority of observed events that are either intrinsically subluminous
or substantially extinguished by dust, leaving a majority subsample
that has observational absolute--magnitude dispersions of less than
$\sigma_{obs}(M_B) \simeq \sigma_{obs}(M_V) \simeq 0.3$ mag.
Correlations between absolute magnitude and one or more
distance--independent SN~Ia or parent--galaxy observables can be used
to further standardize the absolute magnitudes to better than 0.2 mag.
The absolute magnitudes can be calibrated in two independent ways ---
empirically, using Cepheid--based distances to parent galaxies of
SNe~Ia, and physically, by light curve and spectrum fitting.  At
present the empirical and physical calibrations are in agreement at
$M_B \simeq M_V \simeq -19.4$ or $-19.5$. Various ways that have been
used to match Cepheid--calibrated SNe~Ia or physical models to SNe~Ia
that have been observed out in the Hubble flow have given values of
$H_0$ distributed throughout the range 54 to 67 km s$^{-1}$
Mpc$^{-1}$.  Astronomers who want a consensus value of $H_0$ from
SNe~Ia with conservative errors could, for now, use $60 \pm 10$ km
s$^{-1}$ Mpc$^{-1}$.

\vfill\eject

\noindent 1. INTRODUCTION
\medskip 

One way to illustrate the rapid progress in using Type~Ia supernovae
(SNe~Ia) for cosmology is to compare the situation now with that of
six years ago when the article ``Type~Ia Supernovae as Standard
Candles" (Branch \& Tammann 1992) appeared in this series of reviews.
At that time optimism was expressed about checking that light curves
of SNe~Ia are time--dilated as they should be if the universe really
is expanding; about using SNe~Ia to measure galaxy peculiar
velocities; and about using them to determine the cosmic deceleration
--- but hardly any significant results on these matters were at hand.
Now, thanks to the exertions of supernova observers, the time dilation
has been established (Goldhaber et~al 1997; Leibundgut et~al 1996;
Riess et~al 1997a); peculiar motions are beginning to be estimated
(Riess, Press, \& Kirshner 1995b; Hamuy et~al 1996b; Riess et~al
1997b); and above all, estimates of the matter--density parameter
$\Omega_m$ and the cosmological--constant contribution
$\Omega_\Lambda$ are beginning to be made (Perlmutter et~al 1997b,c;
Garnavich et~al 1998).  All of this work entails using SNe~Ia as
precise indicators of {\sl relative} distances in a purely empirical
way.  Dramatic progress is being made but at this time a review
article would be premature.

A competitive measurement of the Hubble constant, on the other hand,
requires {\sl absolute} distances but less precision.  Traditionally
those who have used SNe~Ia to estimate $H_0$ have obtained values that
have been, in the context of the longstanding distance--scale
controversy, low values.  For example, before Cepheid variables in any
SN~Ia parent galaxy had been discovered Branch \& Tammann (1992)
offered $H_0 = 57 \pm 7$ km s$^{-1}$ Mpc$^{-1}$. (The units of $H_0$
will not be repeated.) Within the last six years so much has been said
and done about obtaining the value of $H_0$ from SNe~Ia that this
alone will be the topic of this review.

Most of the literature citations will be from the 1990s.  Many earlier
ones that are now mainly of historical interest can be found in the
review by Branch \& Tammann (1992).  For a recent comprehensive
collection of articles on practically all aspects of SN~Ia research
see {\sl Thermonuclear Supernovae}, edited by Ruiz--Lapuente et~al
(1997), and for a recent collection of articles on various ways to
estimate $H_0$ see {\sl The Extragalactic Distance Scale}, edited by
Livio et~al (1997).

Sections 2 and 3 focus on empirical matters with the former devoted
to the observational properties of SNe~Ia and the latter to the
determination of $H_0$ by means of Cepheid--based calibrations of
SN~Ia absolute magnitudes. Then the physical properties of SNe~Ia are
discussed in Section 4 and the determination of $H_0$ by physical
methods is the subject of Section~5.  Section~6 states the conclusion.

\bigskip
\noindent 2. OBSERVATIONAL PROPERTIES OF SNe Ia
\medskip

Among the observational advances of recent years some of the
highlights have been as follows.

Several SNe~Ia in relatively nearby galaxies have been well
observed. Normal events include SN~1989B in NGC~3627 of the Leo group
of galaxies (Barbon et~al 1990; Wells et~al 1994); SN~1990N in
NGC~4639 of the Virgo complex (Leibundgut et~al 1991a; Jeffery et~al
1992; Mazzali et~al 1993); SN~1992A in NGC~1380 of the Fornax cluster
(Kirshner et~al 1993; Hamuy et~al 1996d); and SN~1994D in NGC~4526 in
Virgo (Richmond et~al 1995; Patat et~al 1996; Meikle et~al 1996; Vacca
\& Leibundgut 1996).  The two most notoriously peculiar events,
SN~1991T in NGC~4527 (Ruiz--Lapuente et~al 1992; Filippenko et~al
1992a, Phillips et~al 1992; Jeffery et~al 1992; Mazzali, Danziger, \&
Turatto 1995) and SN~1991bg in NGC~4374 (Filippenko et~al 1992b;
Leibundgut et~al 1993; Turatto et~al 1996; Mazzali et~al 1997), were
in the Virgo complex.  SN~1986G in NGC~5128 $\equiv$ Centaurus~A
(Phillips et~al 1987; Cristiani et~al 1992) also was peculiar, in the
sense of SN~1991bg but less extreme.  It is noteworthy that some of
these events were discovered closer to their times of explosion than
to their times of maximum light; that ultraviolet spectra of SN~1992A
were obtained by the Supernova INtensive Study (SINS) collaboration
using the {\sl Hubble Space Telescope} {\bf (HST)} (Kirshner et~al
1993); and that infrared spectra of SN~1994D (Meikle et~al 1996) and
several other SNe~Ia (Bowers et~al 1997) have been observed.

\smallskip

(2) Accurate CCD light curves have been measured for dozens of SNe~Ia
out in the Hubble flow ($z > 0.01$) where the recession velocities of
their parent galaxies should be reliable indicators of their relative
distances.  The discovery and photometry of such events have been
accomplished mainly by the Cal\'an--Tololo collaboration (Hamuy et~al
1993, 1995, 1996c; see also Riess 1996).

\smallskip

(3) SNe~Ia at high redshift by supernova standards, $z > 0.3$, are
being discovered in even larger numbers (Perlmutter et al 1997a,b;
Schmidt et al 1997). {\sl Scheduled} discoveries of whole batches of
such events have become routine (which is not to say easy), thus
allowing the opportunity for follow--up spectroscopy and photometry to
be arranged in advance.  The primary purpose of the search for
high--$z$ SNe~Ia is to use them as indicators of relative distances
for the determination of $\Omega_m$ and $\Omega_\Lambda$ but they also
have been used to constrain the ratio of the global and local values
of $H_0$ to be near (Kim et~al 1997; Tripp 1997) but perhaps slightly
less than (Zehavi et~al 1998) unity.

\bigskip
\noindent {\sl 2.1 Homogeneity and Diversity}
\medskip

Much has been written, pro and con, about the observational
homogeneity of SNe~Ia and for the most part the discussion has
followed a natural course.  Before the 1990s and even in Branch \&
Tammann (1992), when most of the mild apparent differences among
SNe~Ia were comparable to the observational errors, the striking
observational homogeneity was emphasized although it already was clear
both photometrically (Phillips et~al 1987) and spectroscopically
(Branch, Drucker, \& Jeffery 1988) that SNe Ia are not strictly
homogeneous.  After the discovery of the obviously deviant SNe~1991T
and 1991bg it became necessary to split SNe~Ia into a majority of
normal events and a minority of peculiar ones.  In the 1990s the
observational data have improved so much that it has become possible
to look seriously at the diversity even among normal SNe Ia.  This has
brought about a natural new emphasis on the diversity --- but it would
be a mistake to lose sight of the fact that even as the new data have
illuminated the diversity they also have confirmed that {\sl normal
SNe~Ia are highly homogeneous}.  When made in the context of using
SNe~Ia to determine $H_0$, statements such as ``SNe~Ia are far from
standard candles'' and ``the concept of a normal SN~Ia is without
merit'' do more to obscure than to illuminate the reality.

It is clear and uncontroversial that events like SN~1991bg and to a
lesser extent SN~1986G are ``weak'' explosions.  They can be readily
distinguished from normal SNe~Ia on the basis of practically {\sl any}
spectroscopic or photometric observable, so although they are very
interesting physically they need have no bearing on the determination
of $H_0$.  Even if these intrinsically red, dim events did not exist
we would still have to deal with red, dim, extinguished events.
Coping with events like SN~1991T is less straightforward but they are
uncommon and do not pose a serious problem for $H_0$. In the rest of
this section we consider the distributions of various observational
properties of SNe~Ia with some emphasis on the extent to which the
normal events are homogeneous and diverse.

\bigskip
\noindent 2.1.1 Spectra: Observational aspects of supernova spectra
and their classification have been reviewed in this series by
Filippenko (1997).  Given a decent spectrum, deciding whether an event
is or is not a Type~Ia is practically always unambiguous (except at
high redshift). Spectra of three normal SNe~Ia and the peculiar
SNe~1991T, 1991bg, and 1986G, all near maximum light, are displayed in
the right panel of Figure~1. (The left panel will be referred to in
Section~4.4.)  Normal SNe~Ia undergo a characteristic spectral
evolution, showing P~Cygni lines of Si~II, Ca~II, S~II, O~I, and Mg~II
prior to and near maximum light, developing blends of P~Cygni
permitted Fe~II lines shortly thereafter, and finally developing
blends of forbidden emission lines of iron and cobalt ions during the
late nebular phase.  The most conspicuous spectroscopic peculiarity of
weak events like SNe~1991bg and 1986G during their photospheric phase
is that they show a broad absorption trough around 4200
\AA\ produced by low--excitation Ti~II lines (Filippenko et
al 1992a).  In SN~1991T, prior to and around maximum light,
high--excitation lines of Fe~III (Ruiz--Lapuente et~al 1992;
Filippenko et~al 1992b) were prominent. By a few weeks after maximum
light the spectra of SN~1991T looked nearly normal.

Looking at line spectra is a good way to study diversity because the
strengths and blueshifts of the lines (in the SN rest frame) are not
affected by extinction or distance.  In terms of the general
appearance of their spectral features most SNe~Ia look very
similar. For a beautiful illustration of what is meant by SN~Ia
spectral homogeneity see Figure 6 of Filippenko (1997).  Branch,
Fisher, \& Nugent (1993) defined what they meant by spectroscopically
normal (like SNe~1981B, 1989B, 1992A, and 1972E) and peculiar (like
SNe~1986G, or 1991bg, or 1991T) and then found that 83 to 89 percent
of the SNe~Ia whose spectra they were able to subclassify were normal.
Subtle differences in the spectra of normal SNe~Ia were nevertheless
evident.  As more SN~Ia spectra have been published (Gomez, Lopez, \&
Sanchez [1996] set a record by presenting spectra of 27 different
SNe~Ia in one paper) the situation has not changed. Most SNe~Ia in the
observational sample are spectroscopically normal in the sense of
Branch et~al (1993).

Spectroscopic diversity can be quantified by measuring blueshifts of
absorption features or monochromatic flux ratios. Branch \&
van~den~Bergh (1993) compiled data on the wavelength of the absorption
minimum near 6100~\AA\ during the photospheric phase of 36 SNe~Ia and
converted to velocity on the assumption that the feature is produced
by blueshifted $\lambda6355$ of Si~II.  It should be noted that Fisher
et~al (1997) suggested that in pre--maximum spectra this absorption
feature may be partially produced by $\lambda6580$ of C~II; if this is
correct then some of the velocities derived on the basis of the Si~II
identification will need to be revised.  In any case the measured
differences in the feature wavelength at a given phase exceed the
observational uncertainties.  Branch \& van~den~Bergh (1993) defined a
parameter $V_{10}(Si)$, the blueshift at ten days after maximum light
(late enough to be unaffected by C~II). SNe~Ia have $8,000 <
V_{10}(Si) < 14,000$ km s$^{-1}$. Weak SNe~Ia tend to have the very
lowest values but there also are definite differences among normal
SNe~Ia.

For a small sample of 12 SNe~Ia Fisher et~al (1995) defined a
parameter $V_R(Ca)$ that is based on the wavelength of the red edge of
the Ca~II H\&K absorption feature at moderately late times and
intended as a measure of the minimum ejection velocity of calcium.
The weak SNe~1991bg and 1986G had exceptionally low $V_R(Ca)$ values,
1600 and 2700 km s$^{-1}$, respectively, while normal SNe~Ia had $4000
< V_R(Ca) < 7300$ km s$^{-1}$.  For a sample of eight SNe~Ia Nugent
et~al (1995b) defined parameters R(Si) based on the relative depths at
maximum light of the Si~II absorption features near 5800 and 6150~\AA\
, and R(Ca) based on the relative heights of the emission peaks on
both sides of the Ca~II H\&K absorption feature.  Again SN~1991bg and
to a lesser extent SN~1986G had extreme values of both parameters and
clear differences were observed among the normal SNe~Ia.

Thus on the basis of the general appearance of their spectra SNe~Ia
can be divided into a majority of normals and a minority of peculiars.
When spectral features are quantified peculiar weak events have
extreme values of the parameters.  Moderate spectral diversity among
normal SNe~Ia also exists and given good data it can be quantified in
various ways.

\bigskip \noindent 2.1.2 Light--Curve Shapes: Like spectra,
rest--frame light--curve shapes are independent of distance and (to
first order) of extinction.  And like SN~Ia spectra, SN~Ia
light--curve shapes are impressively homogeneous (Leibundgut et~al
1991b; Hamuy et~al 1991; Branch \& Tammann 1992).  The first obviously
deviant light curves were those of SN~1986G (Phillips et~al 1987)
which was fast and SN~1991bg (Filippenko et~al 1992a; Leibundgut et~al
1993) which was faster.  Now that good photometry is available for
numerous SNe~Ia more moderate differences among light curves can be
studied.  The parameter that most often has been used to quantify
light--curve shapes is $\Delta m_{15}$ (Phillips 1993), the magnitude
decline in the $B$ band during the first 15 days after maximum light.
Often, when the data for a direct measurement of $\Delta m_{15}$
aren't available, a value is assigned by fitting the data to template
light curves of SNe~Ia having various values of $\Delta m_{15}$.
Hamuy et~al (1996d) presented a set of six templates in the $B, V,$
and $I$ bands ranging from the rather slow light curves of SNe~1992bc
($\Delta m_{15}=0.87$) and 1991T ($\Delta m_{15}=0.94$) to the very
rapid ones of SN~1991bg ($\Delta m_{15}=1.93$).  The SN~1992A light
curves ($\Delta m_{15}=1.47$) are shown in Figure 2 as an example of
superb photometric data and a comparison of the shapes of the six
templates is shown in Figure 3.  Light--curve shape diversity exists
and given good data it can be quantified.

\bigskip \noindent 2.1.3 Colors and Absolute Magnitudes: Colors and
absolute magnitudes are both affected by interstellar extinction so I
discuss them together.  A source of confusion in the past was that a
few SNe~Ia were reported to be very blue at maximum light, with $B - V
\simeq -0.3$.  (Throughout this article $B-V$ will be the difference
between the peak $B$ and peak $V$ values, i.e., it is really
$B_{max}-V_{max}$.)  This implied either a wide range in intrinsic
color among normal SNe~Ia or a characteristic very blue intrinsic
color with considerable interstellar reddening being the norm rather
than the exception.  However, CCD photometry of comparison stars that
were used for the old photographic photometry has confirmed a previous
suspicion that the ``too--blue'' colors were in error (Patat et al
1998) and now the color distribution of SNe~Ia is strongly peaked near
$B-V=0$. The minority of events that have much redder values include
some like SN~1989B ($B-V=0.35$), which were intrinsically normal but
obviously reddened by dust, and weak events like SN~1991bg
($B-V=0.87$), which were intrinsically red. That most SNe~Ia should
escape severe extinction in their parent galaxies is not surprising
(Hatano et~al 1998).  In addition, of course, observational selection
works against the discovery and follow--up of the small fraction of
SNe~Ia that do happen to be severely extinguished.

For the distribution of peak absolute magnitudes, relative distances
are required.  On the basis of a sample that included data from Hamuy
et~al (1995) as well as some earlier less accurate data, Vaughan et~al
(1995) and Tammann \& Sandage (1995) stressed that when a simple $B-V$
criterion is used to exclude SNe~Ia that are observationally red,
whether intrinsically or because of dust, one is left with a sample of
nearly standard candles.  This has been confirmed by Hamuy et~al
(1996b).  The top panel of Figure 4 shows the Hubble diagram in $V$
for the 26 events in their sample having $B-V < 0.2$. (The bottom
panel will be referred to in section 2.2.) The observational
absolute--magnitude dispersions of these events are only
$\sigma_{obs}(M_B)=0.24$, $\sigma_{obs}(M_V)=0.22$, and
$\sigma_{obs}(M_I)=0.19$.  Applying a color cut is a very
straightforward strategy for obtaining a sample of nearly standard
candles, but it can be misunderstood (von Hippel, Bothun, \& Schommer
1997).  The mean absolute magnitudes, corrected for Galactic but not
parent--galaxy extinction, of the 26 events of Hamuy et~al (1996b)
that pass the color cut are

$$ M_B \simeq M_V \simeq -19.30 (\pm 0.03) + 5 \log (H_0/60).
\eqno (1) $$

\noindent Saha et~al (1997) show a Hubble diagram (their Figure 10)
for a larger sample of 56 SNe~Ia that includes some more recent data
(Riess 1996) but also some older data that are not as precise as those
of Hamuy et al (1996b); therefore the dispersions for the Saha et~al
sample are somewhat higher, $\sigma_{obs}(M_B)=0.33$ and
$\sigma_{obs}(M_V)=0.31$.  Wang, H\"oflich, \& Wheeler (1997) point
out that the scatter in the absolute magnitudes, uncorrected for
extinction, of SNe~Ia beyond about 7.5 kpc from the centers of their
parent galaxies is very small; unfortunately none of the
Cepheid--calibrated SNe~Ia to be discussed in section 3 were that far
out.

Figure 5 shows absolute magnitude plotted against $B-V$ for the
Cal\'an--Tololo sample. For the entire sample, including the weak
SN~1992K that resembled SN~1991bg (Hamuy et~al 1995a), the weak 1993H
that probably was like SN~1986G (Hamuy et~al 1996b), and the
moderately extinguished SN~1990Y (Hamuy et~al 1995), a correlation is
obvious, and even when the three red events are excluded a correlation
remains.  In the past the correlation between color and
absolute--magnitude seemed to require that the extinction
characteristics of the dust that reddens SNe~Ia in external galaxies
differ from those of the dust in our Galaxy (Branch \& Tammann 1992
and references therein).  Recently this problem has been thought to
have been due to the old ``too--blue'' SNe~Ia mentioned above and to
the adoption of a single intrinsic value of $B-V$ for SNe~Ia (Della
Valle \& Panagia 1992; Sandage \& Tammann 1993; Vaughan et~al 1995;
Riess et~al 1996b).  Strangely enough the mystery appears to have
resurfaced (Tripp 1998).

\bigskip
\noindent {\sl 2.2 Correlations}
\medskip

Here we consider correlations between observables other than the one
just discussed between color and absolute magnitude.  As stressed
above weak events like SN~1991bg are extreme in many of their
properties so when normal and weak SNe~Ia are considered together
correlations are obvious.  The more interesting issue is the extent to
which correlations hold among normal SNe~Ia.  First consider those
among distance--independent observables.

It is clear that there is a correlation between spectrum and
light--curve shape.  Nugent et al (1995b) found that the spectroscopic
observables R(Ca) and R(Si) correlate well with the light--curve
decline rate $\Delta m_{15}$.  Therefore the concept of a
one--dimensional photometric/spectroscopic sequence is useful.  The
situation is not really that simple though.  For example, some events
whose spectra look normal in the sense of Branch et~al (1993) do not
fit into a one--dimensional sequence in terms of the Si~II blueshifts
studied by Branch \& van den Bergh (1993).  SNe~1983G and 1984A had
normal looking spectra but exceptionally high $V_{10}(Si)$ values.
Similarly Hamuy et al (1996d) find that although in general their
light--curve templates can be arranged in a one--dimensional sequence
from slow to fast, some light curves having very similar initial
decline rates show significant differences of detail. SN 1994D had an
anomalously negative $U-B$ color for its decline rate.  And although
SN~1992bc had a slower light curve than the spectroscopically peculiar
SN~1991T, SN~1992bc had a normal looking spectrum.  This and other
evidence suggests that events like SN~1992bc may represent the
``strong'' end of the sequence of normals, while SN~1991T and similar
events that have been discovered recently comprise a separate subgroup
of powerful SNe~Ia which may be super--Chandrasekhar products of
white--dwarf mergers in young populations (see section 4.4).

Looking for correlations between $B-V$ and other distance--independent
observables is difficult because for all but the weak SNe~Ia the
intrinsic $B - V$ distribution is so strongly peaked near zero.  The
$U-B$ color appears to cover more of a range than $B-V$ (Schaefer
1995b) and to be correlated with other SN~Ia properties (Branch,
Nugent, \& Fisher 1997) but more $U-B$ data are needed.

Spectra and light curves of SNe~Ia also are correlated with the nature
of the parent galaxies.  On average SNe~Ia in red or early--type
galaxies have lower values of $V_{10}(Si)$ and faster light curves
than those in blue or late--type galaxies (Hamuy et~al 1995; Branch,
Romanishin, \& Baron 1996b; Hamuy et~al 1996a).  The $\Delta m_{15}$
parameter is plotted against the color of the parent galaxy in Figure
6.

Correlations with {\sl absolute magnitude}, which we consider next,
are necessary for improving on the standard--candle approach to $H_0$.
Using Tully--Fisher (TF) and surface--brightness--fluctuation (SBF)
distances for a sample of nine SNe~Ia including the peculiar
SNe~1991T, 1986G, and 1991bg, and the not so well observed SN~1971I,
Phillips (1993) found correlations between absolute magnitude ($M_B,
M_V$, and $M_I$) and $\Delta m_{15}$.  At one extreme SN~1991T was
overluminous and somewhat slow to decline and at the other SN~1991bg
was extremely subluminous and quick to decline.  The correlation was
in the same sense as that proposed by Pskovskii (1977).  Subsequently
the Cal\'an--Tololo data on SNe~Ia in the Hubble flow, for which the
relative distances are more secure, showed that the correlation
indicates considerable scatter among the dim, red, rapidly-declining
events. After adopting a color cut to eliminate the three
observationally red events in their sample Hamuy et~al (1996b) derived
slopes $dM_B/d\Delta m_{15} = 0.78\pm0.17$, $dM_V/d\Delta m_{15}=
0.71\pm0.14$, and $dM_I/d\Delta m_{15}=0.58\pm0.13$ (Figure~7).  These
slopes are much less steep than those that were obtained by Phillips
(1993) yet definitely greater than zero.  The Hubble diagram in $V$
for the 26 non--red events, standardized by means of $\Delta m_{15}$,
is shown in the bottom panel of Figure 4.  When the correlations with
$\Delta m_{15}$ were taken into account the absolute--magnitude
dispersions fell from those quoted in section 2.1.3 to
$\sigma_{obs}(M_B)= 0.17$, $\sigma_{obs}(M_V)=0.14$, and
$\sigma_{obs}(M_I)=0.13$.  Tripp (1997) found slopes that agreed well
with those of Hamuy et~al (1996b).  In a straightforward but rigorous
statistical analysis of the Cal\'an--Tololo data Tripp (1998) also
introduced a linear dependence of absolute magnitude on $B-V$ and
found that $dM_B/d\Delta m_{15}$ falls to about 0.5; that $B-V$ is
more effective than $\Delta m_{15}$ in standardizing absolute
magnitudes; and that the two--parameter corrections transform the
Cal\'an--Tololo sample into perfectly standardized candles insofar as
can be measured with present techniques.

Absolute magnitude also correlates with the spectroscopic indices
discussed by Fisher et~al (1995) and Nugent et~al (1995b), with
parent--galaxy type (Hamuy et al 1996a; Saha et~al 1997), and with
parent--galaxy color (Hamuy et~al 1995; Branch et al 1996b). On
average normal SNe~Ia in early--type or red galaxies appear to be
dimmer by 0.2 or 0.3 mag than those in late--type or blue galaxies and
the absolute--magnitude dispersions of non--red SNe~Ia in blue and in
red galaxies, considered separately, are only about 0.2.

In an analysis that bears on many of the issues that have been
discussed in this section, Riess, Press, \& Kirshner (1995a, 1996a)
have developed a formal statistical procedure to simultaneously
estimate extinctions, relative luminosities and relative distances
from the shapes of light curves in the $B, V, R,$ and $I$ bands
(multicolor light--curve shapes: MLCS).  A training set of 9 SNe~Ia
with adopted distance and extinction estimates was used to establish a
linear relationship between luminosities and light--curve shapes.
Then individual extinctions, relative luminosities, and relative
distances were inferred for an independent set of 20 Hubble--flow
SNe~Ia (including 10 from the Cal\'an--Tololo survey), from their
light curves.  An attractive feature of MLCS is that it yields formal
error estimates.  The analysis gave results for various SN~Ia
properties such as relations between $B-V, V-R$, and $R-I$ color
curves and luminosity (Figure 8), and the dispersion in the MLCS
distance moduli was found to be only 0.12 mag.  These SN~Ia properties
are less directly ``observational'' than those discussed earlier in
this section because they depend on the adopted properties of the
training set including TF and SBF relative distances, and on an
assumed {\sl a priori} probability distribution of the extinction.  Of
the 9 training--set events, 8 also were in the sample of 9 that was
used by Phillips (1993) to derive magnitude--decline slopes that
proved to be too steep, and the maximum--light $B-V$ as parameterized
by Riess et~al (1996a) correlates with luminosity (Lira 1996) and
$\Delta m_{15}$ (R. Tripp, private communication) in ways that the
Hubble--flow events of the Cal\'an--Tololo sample do not.  Assessing
the consequences of changing the MLCS input data is not trivial; it
will be interesting to see whether and how the results of MLCS change
when the nearby training set is replaced by a subset of the
Hubble--flow SNe~Ia whose relative distances will be more secure.  See
Riess et~al (1998) for an interesting proposal to circumvent the labor
of obtaining accurate multicolor light--curve shapes by combining the
information carried in a single spectrum and a single--epoch
measurement of $B$ and $V$ magnitudes to determine ``snapshot''
distances to SNe~Ia.

\bigskip
\noindent {\sl 2.3 Summary}
\medskip

SNe~Ia can be divided into a majority of ``normal" events that have
highly homogeneous properties and a minority that are ``peculiar".
When normals and peculiars are considered together correlations among
their observational properties are obvious.  To a certain extent the
correlations hold among normal SNe~Ia and the concept of a sequence of
SNe~Ia ranging from those having high--excitation spectra, high
blueshifts of spectral features, slow light curves, and high
luminosities to those having low--excitation spectra, low blueshifts,
fast light curves and low luminosities, is useful.  The former tend to
occur in blue, late--type galaxies and the latter in red, early--type
galaxies.  A one--dimensional sequence of SNe~Ia is not, however, the
whole story.

A simple $B-V$ cut that eliminates events that are observationally red
at maximum light yields a sample of nearly standard candles having
$\sigma_{obs}(M_B) \simeq \sigma(M_V)_{obs}$ less than 0.3.
Correlations between absolute magnitude and one or more SN~Ia or
parent --galaxy observables can be used to further standardize the
absolute magnitudes to better than 0.2 mag.  As far as determining the
value of $H_0$ by means of Cepheid--calibrated SNe~Ia is concerned,
the accuracy of the result will depend not only on the tightness of
the intrinsic correlation among normal SNe~Ia but also on the degree
to which the relevant distance--independent observables happen to be
known for the Cepheid--calibrated SNe~Ia.

\bigskip
\noindent 3.  $H_0$ FROM CEPHEID CALIBRATIONS
\medskip

Cepheid--based calibrations of the absolute magnitudes of SNe~Ia began
with the discovery and measurement of Cepheid variables in IC~4182,
the parent galaxy of SN~1937C, by the HST SN~Ia Consortium (Sandage
et~al 1992; Saha et~al 1994). The Cepheid distance modulus of IC~4182,
$\mu = 28.36 \pm 0.09$, proved to be in agreement with the modulus
that had been obtained much earlier by Sandage \& Tammann (1982) on
the basis of the brightest red stars in IC~4182, $\mu = 28.21 \pm
0.2$.  As of this writing the Consortium has determined Cepheid
distances to four other galaxies in which five SNe~Ia have appeared
and hopes to be able to determine the distances to at least three more
galaxies including NGC~3627, the parent of SN~1989B; NGC~4527, the
parent of SN~1991T; and NGC~1316$\equiv$Fornax~A, the rather
early--type (Sa peculiar) parent of SNe~1980N and 1981D.  The HST
$H_0$ Key Project team (Freedman et~al 1997) is expected to contribute
a distance to NGC~4414, the parent of SN~1974G.

A critical assessment of the reliability of the Cepheid distances
would be beyond the scope of this article.  Thorough discussions of
the discovery and photometry of the Cepheids, the observed
period--luminosity relations, and the extinction and derived distances
are given in the papers of Saha et~al (1994, 1995, 1996a,b, 1997). The
Consortium uses the $V$ and $I$ band Cepheid period--luminosity
relations of Madore \& Freedman (1991) which are assumed to be
independent of metallicity.  Mild revision of the zero--points of the
$P-L$ relations on the basis of parallaxes of Galactic Cepheids
measured by the {\sl Hipparcos} satellite, in the sense of making
Cepheid distances longer and Cepheid--based estimates of $H_0$ lower,
may prove to be required (Feast \& Catchpole 1997; Madore \& Freedman
1998; Sandage \& Tammann 1998). The metallicity dependence of Cepheid
distances (Gould 1994; Sasselov et~al 1997; Kochanek 1997; Saha et~al
1997; Kennicutt et~al 1998) is an important issue, but the $P--L$
relations in $M_{bol}$ appear to be independent of [Fe/H] at the level
of 0.05 mag for even a change of 1.7 dex in the metalicity (Stothers
1988; Chiosi et~al 1993; Sandage 1996; Saio \& Gautschy 1998).

\bigskip
\noindent {\sl 3.1 The Cepheid--Calibrated SNe~Ia}
\medskip

Table~6 of Saha et~al (1997) lists distance moduli, apparent
magnitudes, and absolute magnitudes for seven Cepheid--calibrated
SNe~Ia: SN~1937C in IC~4182; SNe~1895B and 1972E in NGC~5253; SN~1981B
in NGC~4536; SN~1960F in NGC~4496A; SN~1990N in NGC~4639; and SN~1989B
in NGC~3627.  The Cepheid distance of the last galaxy has not yet been
determined directly; it is assigned the mean Cepheid distance of three
fellow members of the Leo group of galaxies.  There has been little
disagreement about the apparent magnitudes of these calibrator SNe~Ia
except in the case of SN~1937C (Schaefer 1994; Pierce \& Jacoby 1995;
Schaefer 1996a; Jacoby \& Pierce 1996).  Extinction of the calibrators
by dust in our Galaxy is low.  Estimates of the extinction of the
SNe~Ia by dust in the parent galaxies are not so easy to come by but
parent--galaxy extinction of at least four of the calibrators,
SNe~1895B, 1937C, 1960F, and 1972E, is likely to have been quite low.
Saha et~al (1997) assumed that the total extinction of these four was
the same as the (low) mean extinction of the corresponding Cepheids,
in which case the apparent distance modulus of the Cepheids combined
with the apparent magnitude of the SN~Ia gives the extinction--free
SN~Ia absolute magnitude.  For SNe~1981B, 1989B, and 1990N Saha et al
obtained absolute magnitudes using the extinction--corrected distance
moduli of the Cepheids together with assumed supernova extinctions of
$E(B-V)=0.1$, 0.37 (Wells et al 1994), and 0.00, respectively.  The
mean absolute magnitudes listed by Saha et~al are $M_B=-19.52 \pm
0.07$ based on seven events and $M_V = -19.48 \pm 0.07$ based on six
events.

Spectroscopically, SNe~1960F, 1981B, 1989B, and 1990N were observed at
maximum light and they were normal.  The earliest spectra of SNe~1937C
and 1972E were obtained long enough after maximum light to raise the
issue of whether they could have been peculiar in the sense of
SN~1991T; arguments that they were spectroscopically normal have been
made by Branch et~al (1993, 1994).  The single spectrum of SN~1895B
(Schaefer~1995a) is consistent with normal but an SN~1991T--type
peculiarity cannot be excluded.

The inclusion of SN~1895B as a calibrator has been questioned on the
grounds that its apparent magnitude may be unreliable but Saha et~al
(1995) argued that it would be arbitrary to exclude it because there
is no reason to expect a {\sl systematic} error in its apparent
magnitude.  Also, it must be remembered that the uncertainty in a
calibrator's absolute magnitude depends on the uncertainties not only
in the apparent magnitude but also in the extinction and in the
distance modulus; thus the uncertainties in the absolute magnitudes of
the calibrators listed by Saha et~al (1997) turn out to be of
comparable size.  The possibility that SN~1895B was a peculiar event
like SN~1991T may be a better argument for its exclusion. If it is
excluded, and if the most recent (and fainter) estimates of the
apparent magnitudes of SN~1937C (Jacoby \& Pierce 1996) and SN~1990N
(Lira et~al 1998) are adopted, the mean absolute magnitudes of the
seven calibrators only drop from the values of Saha et~al (1997),
$M_B=-19.52$ and $M_V=-19.48$, to $M_B=M_V=-19.44$.

In principle the calibrators could provide an independent test of the
magnitude--decline relation found for the Hubble--flow SNe~Ia but
considering the uncertainties in the absolute magnitudes and the
decline rates of the calibrators the shallow slopes found by Tammann
\& Sandage (1995) and Hamuy et~al (1996a) can be neither confirmed nor
denied.  It is worth noting, though, that invoking a strong
metallicity dependence of the Cepheid zero point in the sense of
reducing the Cepheid distances to metal--deficient galaxies (IC~4182,
NGC 5253) relative to the distances of more metal--rich galaxies would
lead to an {\sl inverse} magnitude--decline relation that is not
plausible.  This is a reason to doubt that introducing a metallicity
dependence will cause any substantial revision of the results
discussed in the next section.

\bigskip
\noindent {\sl 3.2 Matching the Calibrators to the Hubble Flow}
\medskip

The ways that have been used to obtain $H_0$ by comparing the local
Cepheid--calibrated SNe~Ia to Hubble--flow SNe~Ia will be considered
here more or less in order of increasing complexity.  

The intent of the HST SN~Ia Consortium all along has been to simply
treat SNe~Ia as nearly standard candles.  In their most recent paper
(Saha et~al 1997) they assigned mean values for $M_B$ and $M_V$ of the
calibrators to the mean values of 56 non--red ($B-V < 0.2$)
Hubble--flow SN~Ia to obtain $H_0=58^{+7}_{-8}$, where extinction of
the non--red Hubble--flow SN~Ia is neglected.  Saha et~al emphasized
that for $H_0$ higher than 65 nearly all of the Hubble---flow SNe~Ia
would be intrinsically dimmer than nearly all of the
Cepheid--calibrated SNe~Ia (Figure~9) in spite of the observational
bias in favor of discovering the most luminous of the Hubble--flow
SNe~Ia.  The larger the SN~Ia absolute--magnitude dispersion actually
is (cf. van den Bergh 1996), the stronger an argument this is for a
low value of $H_0$ because the effect of the bias increases with the
size of the dispersion.  For an insightful discussion of the effects
of selection bias on distance determinations see Teerikorpi (1997) in
this series.

Others have used the Saha et~al Cepheid distances, together with
slightly different adopted apparent magnitudes of the SN~Ia
calibrators and in some cases using only subsets of the calibrators,
to make their own estimates of $H_0$.  For example, the local
calibrators are in blue galaxies and since SNe~Ia in blue galaxies
tend to be more luminous than those in red galaxies it would seem to
be appropriate to match the calibrators only to Hubble--flow SNe~Ia in
blue galaxies.  Branch et~al (1996b) obtained $H_0=57\pm4$ in this
way, and Saha et~al (1997) show that this hardly affects their result
for $H_0$.  If a Cepheid distance to NGC~1316 can be obtained it will
provide a valuable calibration of two SNe~Ia in an early type, red
galaxy.

Another way to match the calibrators to Hubble--flow SNe~Ia is to use
the supernova color to standardize the absolute magnitudes.  The
correlation between absolute magnitude and $B-V$ will be anything but
perfect because even if a one--to--one intrinsic relation existed it
wouldn't be exactly like the relation produced by
extinction. Nevertheless this approach has the attractive feature that
it requires no extinction estimates for the calibrators {\sl or} the
Hubble--flow events.  In this way Branch et al (1996a), using several
subsets of calibrators (none of which included SN~1895B because there
is no estimate of its $V$ magnitude) and several subsets of
Hubble--flow SNe~Ia, obtained values of $H_0$ ranging from $54\pm5$ to
$60\pm6$.

The standardization method for absolute magnitudes that has received
the most attention so far is the use of light--curve decline rates or
shapes.  Hamuy et al (1996b) matched the calibrators to their non--red
Hubble--flow SNe Ia using their linear relations between absolute
magnitude and $\Delta m_{15}$ and obtained $H_0=63\pm3.4$ (internal)
$\pm2.9$ (external).  The requirement of an accurate value of $\Delta
m_{15}$ restricted the number of calibrators to four: SNe~1937C,
1972E, 1981B, and 1990N, and extinction of the non--red Hubble--flow
SNe~Ia was neglected.  Schaefer (1996b) considered the apparent
magnitudes of the calibrators in the $U, B, V$ and $H$ bands (where
possible) and obtained $H_0 = 55 \pm 3$.  Kim et~al (1997) matched the
calibrators to five of the Perlmutter et~al (1997b) high--redshift
SNe~Ia with the two samples adjusted to a common mean value of $\Delta
m_{15}$ and obtained estimates of $H_0$ that depend slightly on the
values of $\Omega_m$ and $\Omega_\Lambda$ (see their Figure 3).  For
example, for $\Omega_m = 1$ and $\Omega_\Lambda = 0$ their result for
$H_0$ was $59\pm 3$, whereas for $\Omega_m = 0.3$ and $\Omega_\Lambda
= 0$ it was $62\pm4$.  Tripp (1997) matched the calibrators to a
combination of the Cal\'an--Tololo and high--redshift samples, all
adjusted to a common mean $\Delta m_{15}$, and obtained $H_0=60 \pm
5$.  When Tripp (1998) allowed absolute magnitude to depend linearly
on both $B-V$ and $\Delta m_{15}$ he obtained $H_0 = 60 \pm 6$.
Freedman, Madore, \& Kennicutt (1997) assumed that a Cepheid--based
distance to the spiral galaxy NGC~1365 gives the distance to the
Fornax cluster and to the early--type galaxies NGC 1380, the parent of
SN~1992A, and NGC~1316, the parent of SN~1980N; applying a
magnitude--decline relation to these two events together with some of
the other calibrators they obtained $H_0 = 67 \pm 8$.  But assigning
the distance of NGC~1365 to the early--type parent galaxies of
SNe~1992A and 1980N is highly controversial (Tammann \& Federspiel
1997; Freedman 1997; Sandage \& Tammann 1997; Saha et~al 1997).

Riess et~al (1996a) used their MLCS technique with three calibrators,
SNe~1972E, 1981B, and 1990N, to calibrate the absolute distances to
their sample of 20 Hubble--flow SNe~Ia and obtained $H_0=64\pm3$
(statistical) $\pm6$ (total error).  In this approach individual
extinctions of the Hubble--flow events are derived rather than assumed
or neglected.

\bigskip
\noindent {\sl 3.3 Summary}
\medskip

The mild differences between the results for $H_0$ quoted in this
section are due to differences in the numbers of calibrators that are
used and in the ways that they are matched to Hubble--flow SNe~Ia.
Different people would assign different relative weights to these
analyses.  In principle the analyses that try to take diversity into
account are to be preferred but at present some of these can use only
a small numbers of calibrators.  Suffice it to say that present
Cepheid calibrations of SNe~Ia give values of $H_0$ near 60.  Mild
revision is to be expected as more Cepheid--calibrated and
Hubble--flow SNe~Ia become available, as the zero--points of the
Cepheid $P-L$ relations are revised, and as a Cepheid metallicity
dependence is taken into account, if needed.

\bigskip
\noindent 4. PHYSICAL PROPERTIES
\medskip

The immediate progenitors of SNe~Ia are believed to be carbon--oxygen
(C--O) white dwarfs in close binary systems and no other kind of
progenitor has been under serious consideration for some time.  During
the 1980s, general consensus was also that the first nuclear ignition
was of carbon, deep inside the white dwarf. This event would be
followed by the outwards propagation of a subsonic nuclear flame (a
{\sl deflagration}), the velocity of which had to be (and still has to
be) parameterized.  One particular nuclear--hydrodynamical
deflagration model called W7 (Nomoto, Thielemann, \& Yokoi 1984) was
parameterized in such a way that its radial composition structure with
(Branch et~al 1985) or without (Harkness 1991a) {\sl ad hoc}
compositional mixing of its outer layers was able to give a good
account for the spectral features of normal SNe~Ia.  Thus W7 became
the standard SN~Ia model and this is where matters stood when the
physics of supernovae was reviewed in this series by Woosley \& Weaver
(1986) and still at the time of Branch \& Tammann (1992). Since then
many people have put much effort into seeking the nature of the
progenitor binary systems, constructing hydro explosion models, and
calculating light curves and spectra of models.

\bigskip
\noindent {\sl 4.1 Progenitors}
\medskip

Although numerous articles and several recent reviews (Branch et~al
1995, Renzini 1996, Iben 1997, Ruiz--Lapuente et~al 1997a) have been
written about SN~Ia progenitors we still do not know even whether (or
how often) the progenitor binary system contains one white dwarf or
two.

In the standard {\sl single degenerate} scenario the white dwarf
accretes from the Roche lobe or wind of a nondegenerate companion
until it approaches the Chandrasekhar mass and ignites carbon deep in
its interior.  There has been much recent interest in the possibility
that single degenerate pre--SN~Ia systems are being observed as
supersoft X--ray sources (van den Heuvel et~al 1992; Rappaport, Di
Stefano, \& Smith 1994; Yungelson et~al 1996), especially since
Hachisu, Kato, \& Nomoto (1996) found a new strong--wind solution for
mass transfer from a lobe--filling companion.  According to Hachisu
et~al the formation and expulsion of a common--envelope can be more
easily avoided than previously believed.  This may open up two
promising channels (Figure~10) for the accretor to reach the
Chandrasekhar mass and both could be observed as supersofts: close
systems in which the donor is a main sequence or subgiant star in the
range 2 to 3.5 $M_\odot$ and wide systems in which the donor is a red
giant of $\simeq 1 M_\odot$ (Hachisu et~al 1996; Nomoto et~al 1997; Li
\& van den Heuvel 1997, but see Yungelson \& livio 1998).  In the
single--degenerate scenario a significant amount of circumstellar
matter is expected to be in the vicinity of the explosion.  So far no
convincing evidence for narrow circumstellar hydrogen or helium lines
in SN~Ia spectra has been found (Cumming et~al 1996; Ho \& Filippenko
1995) nor has X--ray (Schlegel \& Petre 1993) or radio (Eck et~al
1996) emission from circumstellar interaction been seen.  These
non--detections are not yet quite stringent enough, however, to rule
out single--degenerate progenitor systems (Lundqvist \& Cumming 1997).
So far, only one SN~Ia has been found to be polarized (Wang, Wheeler,
\& H\"oflich 1997); the general lack of polarization may lead to
constraints on the presence of circumstellar matter and the nature of
the progenitor systems (Wang et~al 1996).

In the standard {\sl double degenerate} scenario two white dwarfs
spiral together as a consequence of the emission of gravitational
radiation to form a super--Chandrasekhar merger product.  According to
a population--synthesis study by Tutukov \& Yungelson (1994) mergers
that form within $3 \times 10^8$ yr of star formation have a mean mass
that is greater than $2 M_\odot$ (Figure~11).  Those researchers who
have made recent attempts to model the merging process using SPH
calculations (Mochkovitch \& Livio 1990; Benz et~al 1990; Rasio \&
Shapiro 1995; Mochkovitch et~al 1997) are not uniformly optimistic
about producing SNe~Ia in this way.  On the other hand there are
arguments (section 4.4) that peculiar events like SN~1991T, at least,
may be super--Chandrasekhar merger products.  It is not clear that
there should be a significant amount of circumstellar matter in the
vicinity of a merger SN~Ia but if so it would be carbon and oxygen
rather than hydrogen and helium.  In this regard the detection of
narrow [O I] lines in late time spectra of SN~1937C by Minkowski
(1939) and the possible detection of narrow O I $\lambda8446$ emission
in a very early spectrum of SN~1991T by Ruiz--Lapuente et~al (1992)
are intriguing.

\bigskip
\noindent {\sl 4.2 Explosion Models}
\medskip

In the 1990s a variety of explosion models other than the classical
deflagration have been considered. They can be divided into ``carbon
ignitors'' and ``helium ignitors'' according to whether the first
nuclear ignition is of carbon, deep inside the accretor, or of helium
near the surface.

Among the carbon--ignitor models, deflagrations, ``delayed
detonations'' (Khokhlov, M\"uller, \& H\"oflich 1993; Arnett \& Livne
1994a,b; Woosley \& Weaver 1994b; H\"oflich 1995; Nomoto et al 1997),
and ``pulsating delayed detonations'' (Khokhlov et~al 1993) could be
applicable in the single degenerate case so these have been
constructed for Chandrasekhar--mass ejection.  In the double
degenerate case the classical idea (see Iben [1997] for a review) is
that the merger leads to a super--Chandrasekhar configuration
consisting of a white--dwarf core, a quasi--spherical
pressure--supported envelope, and a low density thick disk. Whether
this flying saucer (Iben \& Tutukov 1984) eventually explodes or
collapses is thought to depend on, among other things, whether carbon
ignites at the core--disk boundary and burns steadily inward to
produce an oxygen--neon--magnesium configuration that will just
collapse to a neutron star.  A different posssibility is that
explosion of one and then the other white dwarf occurs during or even
just before the merger owing to tidal or shear heating (Iben
1997). Shigeyama et~al (1992) and Khokhlov et~al (1993) constructed
some spherically--symmetric explosion models with super--Chandrasekhar
merger products in mind.

In the 1990s there has been much interest in sub--Chandrasekhar
helium--ignitor models, as constructed by Woosley \& Weaver (1994a),
Livne \& Arnett (1995), and H\"oflich \& Khokhkov (1996).  The first
nuclear ignition is near the bottom of a helium layer of about 0.2
$M_\odot$ accumulated on top of a carbon--oxygen white dwarf.  A
prompt detonation propagates outwards through the helium while an
inward non--burning pressure wave compresses and ignites the
underlying C--O (perhaps well off center) and drives a second
detonation outwards through the C--O.  Owing to the difference between
the nuclear kinetics of carbon and helium burning these models have a
composition structure that is fundamentally different from that of
carbon ignitors.  $^4$He burns to $^{12}$C by the slow triple alpha
process and as soon as $^{12}$C is formed it rapidly captures alpha
particles to form $^{56}$Ni, so the original helium layer ends up as a
high--velocity mixture of $^{56}$Ni and leftover $^4$He.  In these
models intermediate--mass elements such as silicon and sulfur,
produced by low--density carbon burning, are ejected in a relatively
narrow range of velocity around 10,000 km s$^{-1}$.

\bigskip
\noindent {\sl 4.3 Light Curves}
\medskip 

Harkness (1991b) first carried out light--curve calculations that took
into account the dependence of the opacity on temperature, density,
and composition, and found that the light curve of model W7 was about
right for normal SNe~Ia.  Machinery developed for calculating
gamma--ray deposition (H\"oflich, Khokhlov, \& M\"uller 1992) and
bolometric and monochromatic light curves (H\"oflich, Khokhlov, \&
M\"uller 1993) led to a major computational effort by H\"oflich \&
Khokhlov (1996) who calculated LTE light curves for 37 explosion
models encompassing each of the kinds mentioned above.  The light
curves and colors of Chandrasekhar--mass carbon--ignitor models depend
mainly on the amount of $^{56}$Ni that is ejected: models with more
$^{56}$Ni are hotter and brighter and, because the opacity increases
with temperature in the range of interest, they have broader light
curves. Carbon--ignitor models can account reasonably well for the
photometric properties of both normal and peculiar weak SNe~Ia
(Wheeler et~al 1995; H\"oflich et~al 1996, 1997) with normal SNe~Ia
requiring $M_{Ni}\simeq0.6\ M_\odot$ (Figure 12) and SN~1991bg
requiring only about $0.1\ M_\odot$.  The differences between the
calculated light curves for different kinds of carbon--ignitor models
that eject similar amounts of $^{56}$Ni are fairly subtle, so deciding
just which kind of explosion model applies to any particular SN~Ia on
the basis of its light curves and colors alone is difficult.

Discriminating between carbon ignitors and helium ignitors is more
straightforward because they have such different compositions.
H\"oflich \& Khokhlov (1996) found their helium--ignitor models to be
inferior to the carbon ignitors in producing the light curves and
colors of a normal bright SN~Ia. The helium ignitors do not work at
all for subluminous SN~Ia like SN~1991bg because the $^{56}$Ni in the
outer layers keeps the photosphere hot and the colors too blue.

\bigskip
\noindent {\sl 4.4 Spectra}
\medskip

Synthetic spectrum calculations which assumed LTE but were otherwise
detailed and self--consistent were carried out by Harkness (1991b,
1991b). He found that the spectra of model W7 closely resembled those
of the normal SNe~Ia 1981B (Figure~13) and concluded that the ejected
$^{56}$Ni mass needs to be in the range $0.5 - 0.8\ M_\odot$ with the
W7 value, $0.6\ M_\odot$, appearing to be optimum.  Recently, detailed
NLTE spectrum calculations have begun to be made (e.g., Baron et~al
1996; Pauldrach et~al 1996). \Ho (1995) compared synthetic spectra of
delayed--detonation models with spectra of SN~1994D and found that
models having $M_{Ni}\simeq0.6$ provided the best fits.  Nugent et al
(1995a, 1997) found that W7 models provided good fits to the spectra
of SNe~1981B, 1992A, and 1994D (Figure 14).  Nugent et al (1995b)
adopted the W7 composition structure and varied just the effective
temperature to generate a sequence of maximum--light spectra that
resembled the sequence of observed maximum--light spectra all the way
from SN~1991bg through SN~1986G and normal SN~Ia to SN~1991T (Figure
1, left panel).  This means that the differences between the
maximum--light spectra of SNe~Ia, like the differences between the
spectra of stars, are mainly due to differences in temperature.  The
root cause of the temperature range in real SNe~Ia presumably is a
range in $M_{Ni}$, and the whole composition structure surely also
varies along the sequence in a way that has yet to be determined.

\Ho et al (1997) and Nugent et al (1997) also calculated spectra of
helium--ignitor models and found that they did not give satisfactory
fits to the spectra of SNe~Ia.  The exploration of helium--ignitor
models was well motivated on physical grounds (Livne 1990; Woosley \&
Weaver 1994a) so the interesting question about them is why do we not
we see them?  It should be acknowledged here that on the basis of
their calculations of late nebular spectra of explosion models Liu,
Jeffery, \& Schultz (1997a,b) favored sub--Chandrasekhar mass ejection
for normal SNe~Ia and even SN~1991T; Ruiz--Lapuente (1996) did not.
The calculation of nebular spectra is hampered by a lack of reliable
atomic data although the situation is improving (Liu et~al 1997c).  In
any case it is not clear why sub--Chandrasekhar carbon--oxygen white
dwarfs that lack the deadly surface helium layer of the
helium--ignitor models should explode.  Another issue with respect to
the nebular phase is that there are indications from light--curve
shapes (Colgate, Fryer, \& Hand 1997; Cappellaro et~al 1998; Milne,
The, \& Leising 1998) that at late times positrons from Co$^{56}$
decay are not completely trapped as usually assumed.

Detailed NLTE spectrum calculations are invaluable for falsifying
hydrodynamical models but since the number of parameterized hydro
models that can be imagined is infinite and the number of spectra that
can be calculated in NLTE is limited by computational complexity, a
much more empirical approach to supernova spectroscopy also is useful.
Fisher et al (1997) used a fast, parameterized spectrum synthesis code
to study a high quality spectrum of the normal SN~Ia 1990N that was
obtained 14 days before maximum light by Leibundgut et~al (1991a).
Fisher et~al (1997) suggested that the absorption observed near $6040
\AA$, which had been attributed to $\lambda6355$ of Si~II, actually
was produced by $\lambda6580$ of C~II in a high--velocity ($v>26,000$
km s$^{-1}$) carbon--rich region.  Such a layer would be consistent
with published delayed--detonation models.  A. Fisher et al
(manuscript in preparation) suggest that in the peculiar SN~1991T the
``Si~II'' absorption is dominated by C~II $\lambda6580$ before and
perhaps even at maximum light.  On the basis of the empirical
constraints on the composition structure of SN~1991T together with
estimates of the luminosity of SN~1991T, which must be checked with a
Cepheid distance, Branch (1998) and Fisher et~al (in preparation)
suggest that peculiar events like SN~1991T are superluminous, usually
extinguished, substantially super--\Chandra mergers from the youngest
populations that are able to produce SNe~Ia in this way, $\sim 10^8$
yr (Tutukov \& Yungelson 1994).

\bigskip
\noindent {\sl 4.5 Summary}
\medskip

SNe~Ia appear to be carbon ignitors.  If normal SNe~Ia are
Chandrasekhar--mass explosions in single--degenerate progenitor
systems their light curves and spectra indicate that they eject
$M_{Ni}\simeq 0.6\  M_\odot$.  If they are mergers of double degenerates
they should be at least mildly super--Chandrasekhar and then should
eject somewhat more $^{56}$Ni to achieve the temperature needed to
account for their spectra.  

\bigskip
\noindent 5.  $H_0$ FROM PHYSICAL CONSIDERATIONS
\medskip

To first order there is a natural SN~Ia peak luminosity --- the
instantaneous radioactivity luminosity, i.e., the rate at which energy
is being released by $^{56}$Ni and $^{56}$Co decay at the time of
maximum light (Arnett 1982; Arnett, Branch, \& Wheeler 1985; Branch
1992).  With certain simplifying assumptions the peak luminosity is
predicted to be identical to the instantaneous radioactivity
luminosity.  The extent to which they differ, for a hydrodynamical
explosion model, can be determined only by means of detailed light
curve calculations which take into account the dependence of the
opacity on the composition and the physical conditions.  The state of
the art is represented by the calculations of H\"oflich \& Khokhlov
(1996).  The calculated peak luminosity of the models turns out to be
proportional to $M_{Ni}$ within uncertainties, and for models that can
be considered to be in the running as representations of normal SNe~Ia
(carbon ignitors that take longer than 15 days to reach maximum light
in the $V$ band) the characteristic ratio of the peak luminosity to
the radioactivity luminosity is about 1.2 (Branch et~al 1997).  The
physical reason that the ratio exceeds unity in such models was
explained by Khokhlov et~al (1993) in terms of the dependence of the
opacity on the temperature, which is falling around the time of
maximum light.

The H\"oflich \& Khokhlov (1996) light--curve calculations can be used
to estimate $H_0$ in various ways.  H\"oflich \& Khokhlov themselves
compared the observed light curves of 26 SNe~Ia (9 in galaxies having
radial velocities greater than 3000 km s$^{-1}$) to their calculated
light curves in two or more broad bands, to determine the acceptable
model(s), the extinction, and the distance for each event.  From the
distances and the parent--galaxy radial velocities they obtained
$H_0=67\pm9$.  Like the empirical MLCS method, this approach has the
attractive feature of deriving individual extinctions.  But
identifying the best model(s) for a supernova while simultaneously
extracting its extinction and distance, all from the shapes of its
light curves, is a tall order.  This requires not only accurate
calculations but also accurate light curves, and the photometry of
some of the SN~Ia that were used by H\"oflich \& Khokhlov (1996) has
since been revised (Patat et al 1998).  And because the H\"oflich \&
Khokhlov models include many more underluminous than overluminous ones
(the former being of interest in connection with weak SN~Ia like
SN~1991bg) while the formal light--curve fitting technique has a
finite ``model resolution'', a bias towards deriving low luminosities
and short distances for the observed SNe~Ia is possible.

There are less ambitious but perhaps safer ways to use H\"oflich \&
Khokhlov's (1996) models to estimate $H_0$ that involve an appeal to
the homogeneity of normal SNe~Ia and rely only on the epoch of maximum
light when both the models and the data are at their best.  The 10
Chandrasekhar--mass models having $0.49 \le M_{Ni} \le 0.67$ --- i.e.,
those having $M_{Ni}$ within the acceptable range for normal SNe~Ia
--- have a mean $M_{Ni}=0.58$ \msun\ and a mean
$M_V=-19.44$. Alternatively, the five models (W7, N32, M35, M36, PDD3)
that H\"oflich \& Khokhlov found to be most often acceptable for
observed SNe~Ia have a mean $M_{Ni}=0.58$ and a mean $M_V = -19.50$.
Using $M_V=-19.45 \pm 0.2$ in equation (1) gives $H_0= 56 \pm 5$,
neglecting extinction of the non--red Hubble--flow SNe~Ia.

H\"oflich \& Khokhlov's (1996) light--curve calculations were used in
another way by van den Bergh (1995).  Noting that the maximum light
$M_V$ and $B-V$ values of the models obey a relation that mimics that
which would be produced by extinction (Figure 15), he matched the
model relation between $M_V$ and $B-V$ to the relation of the observed
Hubble--flow SN~Ia and obtained values of $H_0$ in the range 55 to 60
depending on how the models were weighted.  If helium--ignitor models
had been excluded, the resulting $H_0$ would have been a bit lower
because the models are underluminous for their $B-V$ colors.  This
procedure has the virtue of needing no estimates of extinction.  The
same result can be seen from Figure 3 of \ho et al (1996) who plot
$M_V$ versus $B-V$ for the H\"oflich \& Khokhlov models and for the
Cal\'an--Tololo SNe~Ia.  For the assumed value of $H_0=65$ the
relations between $M_V$ and $B-V$ for the models and the observed
SNe~Ia are offset and $H_0$ would need to be lowered to about 55 to
bring them into agreement.
  
In their Figure 2 \ho et al (1996) also plot $M_V$ versus a V-band
light--curve decline parameter that is analagous to $\Delta m_{15}$,
for the models and for the Cal\'an--Tololo SNe~Ia.  Again with $H_0=65$
the models and the observed SNe~I are offset with $H_0$ needing to be
lowered to about 55 to bring them into agreement.

Distances to SN~Ia also can be derived by fitting detailed NLTE
synthetic spectra to observed spectra. Nugent et~al (1995c) used the
fact that the peak luminosities inferred from radioactivity--powered
light curves and from spectrum fitting depend on the rise time in
opposite ways, in order to simultaneously derive the characteristic
rise time and luminosity of normal SNe~Ia and obtained $H_0 =
60^{+14}_{-11}$. If SN~Ia atmospheres were not powered by a
time--dependent energy source, the spectrum fitting technique could be
independent of hydrodynamical models.  The procedure would be to look
for a model atmosphere that accounts for the observed spectra without
worrying about how that atmosphere was produced, estimate (or derive
by fitting two or more phases) the time since explosion, and obtain
the luminosity of the model.  But, owing to the time--dependent nature
of the deposition of radioactivity energy, SNe~Ia ``remember" their
history (Eastman 1997; Pinto 1997; Nugent et~al 1997).  Light--curve
and spectrum calculations really are coupled, and more elaborate
physical modeling needs to be done.

\bigskip
\noindent 6. CONCLUSION
\medskip

The mean absolute magnitudes of the SNe~Ia that have been calibrated
by Cepheids in their parent galaxies are $M_B \simeq M_V \simeq -19.4$
or $-19.5$.  The same mean absolute magnitudes are calculated for
explosion models that give a good account of SN~Ia light curves and
spectra, i.e., Chandrasekhar--mass carbon ignitor models that eject
about $0.6\ M_\odot$ of $^{56}$Ni.  Using $M_B= -19.45 \pm 0.2$ in
equation (1) based on the 26 non--red Cal\'an--Tololo Hubble--flow
events would give $H_0 = 56 \pm 5$, but various workers using the
Cepheid--based SN~Ia absolute magnitudes and physical models in
various ways have obtained values of $H_0$ ranging from about 54 to
67.  Those who want a consensus value from SNe~Ia with conservative
errors could use, for now, $H_0 = 60 \pm 10$.

This value depends heavily on three things: the Cepheid distances of
the HST SN~Ia Consortium; the light--curve calculations of H\"oflich
\& Khokhlov (1996); and the observed light curves of the
Cal\'an--Tololo Hubble--flow SNe~Ia.  At present the main issue about
the Cepheid distances appears to be the possibility of a signifcant
metallicity dependence.  Some technical issues that bear on the
accuracy of the complex light--curve calculations remain to be
resolved but the main issue here is the value of the ejected nickel
mass, which in turn depends on whether the total ejected mass is
Chandrasekhar, super, or sub. The main issue about the Hubble--flow
SNe~Ia probably has to do with selection bias.  For example, many of
the Cal\'an--Tololo events peaked not far above the detection limit.
If the Calan--Tololo events are selected from the bright end of the
non--red SN~Ia absolute--magnitude distribution, then the
absolute--magnitude dispersions from the Calan--Tololo survey are
probably too low, the mean absolute magnitude of non--red SNe~Ia in
equation (1) is too bright (for a fixed $H_0$), and most of our
current estimates of $H_0$ from SNe~Ia probably are too high.

This particular reviewer, chock full of opinions and suspicions about
where the errors and biases in the present analyses are likely to be,
goes out on a limb (where a branch belongs) and suggests that when all
has been said and done, $H_0$ from SNe~Ia will be in the 50s.

\bigskip
\bigskip
\noindent ACKNOWLEDGEMENTS
\medskip

The names of colleagues from whom I have learned about SNe Ia range
from Abi to Zalman and would a fill a page, but for numerous
discussions at the University of Oklahoma I must thank Eddie Baron,
Adam Fisher, and Peter Nugent.  Part of this review was written in
Trieste during a stimulating visit to the International School of
Advanced Studies (SISSA) for which I am indebted to my host, Dennis
Sciama.  I want to conclude by expressing my appreciation of Allan
Sandage and Gustav Tammann for their insights and their perseverance.

\vfill\eject

\noindent FIGURE LEGENDS
\bigskip 

\noindent {\sl Figure 1}\ \ NLTE spectra calculated for the
composition structure of model W7 15 days after explosion and a range
of effective temperatures ({\sl left}), and observed spectra of SNe~Ia
near maximum light ({\sl right}).  From Nugent et al (1995b).

\bigskip

\noindent {\sl Figure 2}\ \ Light curves of SN~1992A.  From
Hamuy et al (1996d).

\bigskip

\noindent {\sl Figure 3}\ \ Comparison of six light--curve templates.
From Hamuy et al (1996d).

\bigskip

\noindent {\sl Figure 4}\ \ The Hubble diagram in $V$ for the 26 SNe~Ia in
the Cal\'an--Tololo sample having $B-V \le 0.20$ ({\sl top}), and after
correction for a magnitude--decline correlation ({\sl bottom}).
From Hamuy et al (1996b).

\bigskip

\noindent {\sl Figure 5}\ \ Absolute magnitudes of the Cal\'an--Tololo
sample are plotted against $B-V$.  The extinction vectors have
conventional slopes of 4.1 in B, 3.1 in V, and 1.85 in I.  From Hamuy
et al (1996b).

\bigskip

\noindent {\sl Figure 6}\ \ Light--curve decline rate $\Delta m_{15}$
is plotted against parent galaxy color, corrected to face--on
inclination.  From Branch, Romanishin, \& Baron (1996b).

\bigskip

\noindent {\sl Figure 7}\ \ Absolute magnitudes of the Cal\'an--Tololo
SNe~Ia are plotted against light--curve decline parameter $\Delta
m_{15}$, with weighted linear least--square fits that exclude the
observationally red SNe 1990Y, 1993H, and 1993K.  From Hamuy et al
(1996a).

\bigskip

\noindent {\sl Figure 8}\ \ An empirical family of SN~Ia light and
color curves parameterized by luminosity.  Data for some training--set
members are shown as open circles (SN~1991T); open squares (1994ae);
triangles (1980N); plus signs (SN~1992A); crosses (1986G); and open
diamonds (1991bg). The lines are the light and color curves
constructed from the training--set data, parameterized in terms of
relative absolute magnitude ($\Delta$).  From Riess et~al (1996a).

\bigskip

\noindent {\sl Figure 9}\ \ Absolute magnitude is plotted against
distance modulus for Cepheid--calibrated SNe~Ia (diamonds) and of
other non--red SNe~Ia after 1985 (circles).  The absolute magnitudes
of the latter are shown for three values of $H_0$. From Saha et~al
(1997).

\bigskip

\noindent {\sl Figure 10}\ \ Donor masses are plotted against orbital
period for candidate single--degenerate SN~Ia progenitor binary
systems.  Filled circles are for an initial white--dwarf accretor of
1.2 $M_\odot$ and open circles are for 1.0 $M_\odot$.  The dotted
lines represent the boundaries of mass transfer in case A (left) and
case B (right). From Li
\& van den Heuvel (1997).

\bigskip

\noindent {\sl Figure 11}\ \ The distributions of the masses of
double--degenerate mergers for three age intervals.  The peaks near
2.0, 0.8, and 0.5 $M_\odot$ correspond to CO--CO, CO--He, and He--He
white--dwarf mergers.  Explosions are not expected for total masses
below 1.4 $M_\odot$.  From Tutukov \& Yungelson (1994).

\bigskip

\noindent {\sl Figure 12}\ \ Observed light curves of SN~1994D are
compared with calculated light curves for model M36, a
delayed--detonation that ejected $M_{Ni}=0.6\  M_\odot$ and reached
$M_V= -19.4$.  From H\"oflich (1995).

\bigskip

\noindent {\sl Figure 13}\ \ LTE spectra calculated for model W7 14
days after explosion, with no mixing ({\sl top}) and mixing for $v >
11,000$ km s$^{-1}$ ({\sl center}), are compared with the
maximum--light spectrum of SN~1981B ({\sl bottom}).  From Harkness
(1991a).

\bigskip

\noindent {\sl Figure 14}\ \ NLTE spectra calculated for model W7 are
compared to spectra of SN~1994D at maximum light and SN~1992A 5 days
after maximum light.  From Nugent et al (1997).

\bigskip

\noindent {\sl Figure 15}\ \ $M_V$ is plotted against
$B-V$ for the models of H\"oflich \& Khokhlov (1996).  Helium--ignitor
models are indicated by vertical lines.  The arrow has the
conventional extinction slope of 3.1.  Adapted from van den Bergh
(1996).

\vfill\eject

\noindent {\sl Literature Cited} 
\bigskip

\noindent Arnett WD. 1982.  Type I supernovae. I. Analytic solutions
for the early part of the light curve. {\sl Ap. J.} 253: 785--97

\medskip

\noindent Arnett WD, Branch D, Wheeler JC. 1985.  Hubble's constant
and exploding carbon--oxygen white dwarf models for Type Ia
supernovae.  {\sl Nature} 314: 337--8

\medskip

\noindent Arnett WD, Livne E. 1994a.  The delayed detonation model of
Type Ia supernovae. I. The deflagration phase.  {\sl Ap. J.} 427: 315--29

\medskip

\noindent Arnett WD, Livne E. 1994b.  The delayed detonation model of
Type Ia supernovae.  II.  The detonation phase.  {\sl Ap. J.} 427: 330--41

\medskip

\noindent Barbon R, Benetti S, Cappellaro E, Rosino L, Turatto
		  M. 1990. Type Ia supernova 1989B in NGC 3627.
{\sl Astron. Astrophys.} 237: 79--90

\medskip

\noindent Baron E, Hauschildt PH, Nugent P, Branch D. 1996. Non--local
thermodynamic equilibrium effects in modeling of supernovae near
maximum light.  {\sl MNRAS} 283: 297--315

\medskip

\noindent Benz W, Cameron AGW, Bowers RL, Press WH.  1990. Dynamic
mass exchange in doubly degenerate binaries.  I.  0.9 and 1.2
M$_\odot$ stars.  {\sl Ap. J.}  348: 647--67

\medskip

\noindent Bowers EJC, Meikle WPS, Geballe TR, Walton NA, Pinto PA,
et al. 1997.  Infrared and optical spectroscopy of Type Ia supernovae
in the nebular phase.  {\sl MNRAS} 290: 663--679

\medskip

\noindent Branch D. 1992. The Hubble constant from nickel
radioactivity in Type Ia supernovae.  {\sl Ap J.} 392: 35--40

\medskip

\noindent Branch D. 1998.  Early--time spectra of Type Ia supernovae
and the nature of the peculiar SN~1991T.  In {\sl Supernovae and
Cosmology}, ed. L Labhardt, B Bingelli, R Buser, pp. 000--000.  Schaub
Druck: Sissach, in press

\medskip

\noindent Branch D, Doggett JB, Nomoto K, Thielemann F--K. 1985.
Accreting white dwarf models for Type I supernovae.  IV.  The optical
spectra of carbon deflagration supernovae.  {\sl Ap. J.} 294: 619--25

\medskip

\noindent Branch D, Drucker W, Jeffery DJ. 1988.  Differences among
expansion velocities of Type Ia supernovae.  {\sl Ap. J. Lett.} 330:
L117--18

\medskip

\noindent Branch D, Fisher A, Baron E, \& Nugent P. 1996a.  On van den
Bergh's method for measuring the Hubble constant from Type Ia
supernovae.  {\sl Ap. J. Lett.} 470: L7--9

\medskip

\noindent Branch D, Fisher A, Herczeg TH, Miller DL, Nugent P.
		  1994.  The distance to the Type Ia supernova 1972E
and its parent galaxy NGC~5253: a prediction.  {\sl Ap. J. Lett.} 421:
L87--90

\medskip

\noindent Branch D, Fisher A, Nugent P. 1993.  On the relative
frequencies of spectroscopically normal and peculiar Type Ia
supernovae.  {\sl Astron. J.} 106: 2383--91

\medskip

\noindent Branch D, Livio M, Yungelson LR, Boffi FR, Baron E.
		  1995. In search of the progenitors of Type Ia
supernovae.  {\sl Publ. Astron. Soc. Pac.} 107: 1019--29

\medskip

\noindent Branch D, Nugent P, Fisher A. 1997.  Type Ia supernovae as 
extragalactic distance indicators. See Ruiz--Lapuente, Canal, Isern
1997, pp. 715--34.

\medskip

\noindent Branch D, Romanishin W, Baron E.  1996b.  Statistical
connections between the properties of Type Ia supernovae and the $B-V$
colors of their parent galaxies, and the value of $H_0$.  {\sl Ap. J.}
465: 73--8

\medskip

\noindent Branch D, Tammann GA.  1992.  Type Ia supernovae as standard
candles. {\sl Annu. Rev. Astron.  Astrophys.} 30: 359--89 

\medskip

\noindent Branch D, van den Bergh S. 1993.  Spectroscopic differences
between supernovae of Type Ia in early--type and in late--type
galaxies.  {\sl Astron. J.} 105: 2231--5

\medskip

\noindent Cappellaro E, Mazzali PA,, Benetti S, Danziger IJ, Turatto
		  M, et al. 1998.  SN Ia light curves and radioactive
decay.  {\sl Astron. Astrophys.} 000: 000

\medskip

\noindent Chiosi C, Wood PR, Capitanio N. 1993.  Theoretical Models of
Cepheid Varibales and Their Colors and Magnitudes.  Ap. J. Suppl. 86:
541--98

\medskip

\noindent Colgate SA, Fryer CL, Hand KP. Low mass SNe~Ia and the late
light curve.  See Ruiz--Lapuente, Canal, Isern 1997, pp. 273--302

\medskip

\noindent Cristiani S, Cappellaro E, Turatto M, Bergeron J, Bues
		  I, et al.  1992. The SN 1986G in Centaurus A.
{\sl Astron. Astrophys.} 259: 63--70

\medskip

\noindent Cumming RJ, Lundqvist P, Smith LJ, Pettini M, King
DL. 1996.  Circumstellar H$\alpha$ from SN 1994D and future Type Ia
supernovae: an observational test of progenitor models.  {\sl MNRAS} 283:
1355--60

\medskip

\noindent Della Valle M, Panagia N.  1992.  Type Ia supernovae in late
type galaxies: reddening correction, scale height, and absolute
maximum magnitude.  {\sl Astron. J.} 104: 696--703

\medskip

\noindent Eastman, RG. 1997.  Radiation transport in Type Ia
supernovae.  See Ruiz--Lapuente, Canal, Isern 1997,
		  pp. 571--88

\medskip

\noindent Eck C., Cowan JJ, Roberts D, Boffi FR, Branch D.  1996.
Radio observations of the Type Ia supernova 1986G as a test of a
symbiotic--star progenitor.  {\sl Ap. J. Lett.} 451: L53--5

\medskip

\noindent Feast, MW, Catchpole, RM. 1997. The Cepheid
period--luminosity zero--point from Hipparcos trigonometrical
parallaxes.  {\sl MNRAS} 286: L1-5

\medskip

\noindent Filippenko AV.  1997.  Optical Spectra of Supernovae.  {\sl
Annu. Rev. Astron. Astrophys.}  35: 309-55

\medskip

\noindent Filippenko AV, Richmond MW, Branch D, Gaskell CM,
		  Herbst W, et al. 1992a.  The subluminous,
spectroscopically peculiar Type Ia supernova 1991bg in the elliptical
galaxy NGC 4374.  {\sl Astron. J.} 104: 1543--55

\medskip

\noindent Filippenko AV, Richmond MW, Matheson T, Shields JC,
		  Burbidge EM, et al.  1992b.  The peculiar Type Ia SN
1991T: detonation of a white dwarf?  {\sl Ap J.} 384: L15--8

\medskip

\noindent Fisher A, Branch D, H\"oflich PA, Khokhlov
		  A. 1995. The minimum ejection velocity of calcium in
Type Ia supernovae and the value of the Hubble constant.
{\sl Ap. J. Lett.} 447: L73--6

\medskip

\noindent Fisher A, Branch D, Nugent P, Baron E. 1997.  Evidence for
a high--velocity carbon--rich layer in the Type Ia SN 1990N.
		  {\sl Ap. J. Lett.} 481: L89--92

\medskip

\noindent Freedman WL. 1997.  Determination of the Hubble constant.  In {\sl Critical Dialogues in
Cosmology}, ed. Turok N, pp. 92--129.  Singapore: World Scientific

\medskip

\noindent Freedman WL, Madore BF, Kennicutt RC.  1997.  Hubble Space
Telescope key project on the extragalactic distance scale.  See Livio,
Donahue, Panagia 1997, p. 171--85

\medskip

\noindent Garnavich PM, Kirshner RP, Challis P, Tonry J, Gilliland RL,
et al.  1998.  Constraints on cosmological models from Hubble Space
Telescope observations of high--$z$ supernovae.  {\sl Ap. J. Lett.}
000: 000--

\medskip

\noindent Goldhaber G, Deustra S, Gabi S, Groom DE, Hook I, et
al 1997.  Observations of cosmological time dilation using Type Ia
supernovae as clocks.  See Ruiz--Lapuente, Canal, Isern 1997,
pp. 777--84

\medskip

\noindent G\'omez G, L\'opez R, S\'anchez F.  1996.  The Canarias Type
Ia supernova archive.  {\sl Astron. J.} 112: 2094--109

\medskip

\noindent Gould A. 1994. The metallicity dependence of inferred
Cepheid distances.  {\sl Ap. J.} 426: 542--52

\medskip

\noindent Hachisu I, Kato M, Nomoto K. 1996. A new model for
progenitor systems of Type Ia supernovae.  {\sl Ap. J. Lett.} 470:
L97--100

\medskip

\noindent Hamuy M, Maza J, Phillips MM, Suntzeff NB, Wischnjewsky
		  M, et al. 1993.  The 1990 Cal\'an/Tololo supernova
search. {\sl Astron. J.} 106: 2392--2407

\medskip

\noindent Hamuy M, Phillips MM, Maza J, Suntzeff NB, Della Valle M, et
		  al. 1995a.  SN 1992K: a twin to the subluminous Type
Ia SN 1991bg.  {\sl Astron. J.} 108: 2226--32

\medskip

\noindent Hamuy M, Phillips MM, Maza J, Wischnjewsky M, Uomoto, A
		  et al. 1991.  The optical light curves of SN 1980N
and SN 1981D in NGC 1316 (Fornax A).  {\sl Astron. J.} 102: 208--17

\medskip

\noindent Hamuy M, Phillips MM, Schommer RA, Suntzeff NB, Maza J,
		  Avil\'es R.  1996a.  The absolute luminosities of
the Cal\'an/Tololo Type Ia supernovae.  {\sl Astron. J.} 112: 2391--7
		  
\medskip 

\noindent Hamuy M, Phillips MM, Suntzeff NB, Schommer RA, Maza J,
Avil\'es R.  1995b.  A Hubble diagram of distant supernovae.  {\sl
Astron. J.} 109: 1--13

\medskip

\noindent Hamuy M, Phillips MM, Suntzeff NB, Schommer RA, Maza J,
Avil\'es R. 1996b.  The Hubble diagram of the Cal\'an/Tololo Type Ia
supernovae and the value of $H_0$.  {\sl Astron. J.} 112: 2398--407

\medskip

\noindent Hamuy M, Phillips MM, Suntzeff NB, Schommer RA, Maza J, et
al. 1996c.  $BVRI$ light curves for 29 Type Ia supernovae.
{\sl Astron. J.} 112: 2408--37

\medskip

\noindent Hamuy M, Phillips MM, Suntzeff NB, Schommer RA, Maza J,
		  Smith RC, et al. 1996d.  The morphology of Type Ia
supernova light curves.  {\sl Astron. J.} 112: 2438--47
		  
\medskip

\noindent Harkness R. 1991a.  A comparison of carbon deflagration
models for SNe~Ia.  In {\sl Supernovae}, ed. Woosley, SE, pp. 454--63.
		  New York: Springer

\medskip

\noindent Harkness R. 1991b.  Type Ia supernovae.  In {\sl SN 1987A and Other Supernovae},
		  ed. Danziger, IJ, Kj\"ar, K, pp. 447--56.
		  Dordrecht: Kluwer
\medskip

\noindent Hatano K, Deaton J, Branch D. 1998.  {\sl Ap. J. Lett.} 
000: 000--

\medskip

\noindent Ho LC, Filipppenko AV.  1995.  Probing the interstellar
medium along the lines of site to supernovae 1994D and 1994I.  {\sl
Ap. J.} 444: 165-74; erratum: {\sl 463}: 818

\medskip

\noindent H\"oflich P. 1995.  Analysis of the Type Ia supernova
1994D.  {\sl Ap. J.} 443: 89--108

\medskip

\noindent H\"oflich P., Khokhlov A. 1996.  Explosion models for Type
Ia supernovae: a comparison with observed light curves, distances,
$H_0$, and $q_0$.  {\sl Ap. J.} 457: 500--28 

\medskip

\noindent H\"oflich P, Khokhlov A, \& M\"uller E.  1992.  Gamma--ray
light curves and spectra of Type Ia supernovae.  {\sl
Astron. Astrophys.} 259: 549--66

\medskip

\noindent H\"oflich P, Khokhlov A, \& M\"uller E.  1993.  Light curve
models for Type~Ia supernovae: physical assumptions, their influence
and validity.  {\sl Astron. Astrophys.} 268: 570-90

\medskip

\noindent H\"oflich P., Khokhlov A., Wheeler JC, Nomoto K,
		  Thielemann F--K. 1997. Explosion models, light
curves, spectra, and $H_0$. See Ruiz--Lapuente, Canal, Isern, 1997,
pp. 659--79

\medskip
\noindent H\"oflich P, Khokhlov A, Wheeler JC, Phillips MM,
		  Suntzeff NB, Hamuy M. 1996.  Maximum brightness and
postmaximum decline of light curves of Type Ia supernovae: a
comparison of theory and observations.  {\sl Ap. J.} 472: L81--4

\medskip

\noindent Iben II. 1997. Scenarios for Type Ia supernovae.  See Ruiz--Lapuente, Canal, Isern,
		  1997. pp. 111--26

\medskip

\noindent Iben II, Tutukov AV. 1984.  Supernovae of Type I as end
products of the evolution of binaries with components of moderate
initial mass ($M \le 9\ M_\odot$).  {\sl Ap. J. Suppl.} 54: 335--72

\medskip

\noindent Jacoby GH, Pierce MJ.  1996.  Response to Schaefer's
comments on Pierce \& Jacoby (1995) regarding the Type Ia supernova
1937C.  {\sl Astron. J.} 112: 723--31

\medskip

\noindent Jeffery DJ, Leibundgut B, Kirshner RP, Benetti S,
		  Branch D, Sonneborn G.  1992.  Analysis of the
photospheric epoch spectra of Type Ia supernovae 1990N and 1991T
{\sl Ap. J.} 397: 304--28

\medskip

\noindent Kennicutt RC, Stetson PB, Saha A, Kelson D, Rawson D, et
al. 1998. The {\sl HST} key project on the extragalactic distance
scale XIII.  The metallicity dpendence of the Cepheid distance scale.
{\sl Ap. J.} 000: 000--

\medskip

\noindent Khokhlov A, M\"uller E, H\"oflich
		  P. 1993. Light curves of Type Ia supernova models
with different explosion mechanisms. {\sl Astron. Astrophys.} 270:
223--248

\medskip

\noindent Kim AG, Gabi S, Goldhaber G, Groom DE, Hook IM, et al.
		  1997. Implications for the Hubble constant from the
first seven supernovae at $z \ge 0.35$.  {\sl Ap. J. Lett.} 476:
L63--6

\medskip

\noindent Kirshner RP, Jeffery DJ, Leibundgut B., Sonneborn G,
		  Phillips MM, et al.  1993.  SN 1992A: ultraviolet
and optical observations based on HST, IUE, and CTIO observations.
{\sl Ap. J.} 415: 589--615

\medskip

\noindent Kochanek C. S. 1997.  Rebuilding the Cepheid distance scale.
I. A global analysis of Cepheid mean magnitudes.  {\sl Ap. J.} 491:
13--28

\medskip

\noindent Leibundgut B, Kirshner RP, Filippenko AV, Shields JC,
		  Foltz CB, et al.  1991a.  Premaxium observations of
the Type Ia SN 1990N.  {\sl Ap. J. Lett.} 371: L23--6

\medskip

\noindent Leibundgut B, Kirshner RP, Phillips MM, Wells LA,
		  Suntzeff NB, et al. 1993.  SN 1991bg: a Type Ia
supernova with a difference.  {\sl Astron. J.} 105: 301--13
		  
\medskip

\noindent Leibundgut B, Schommer R, Phillips MM, Riess A, Schmidt
		  B, et al.  1996.  Time dilation in the light curve
of the distant Type Ia supernova 1995K.  {\sl Ap. J. Lett.} 466:
L21--4

\medskip

\noindent Leibundgut B, Tammann GA, Cadonau R, Cerrito D. 1991b.
Supernova studies.  VII.  An atlas of the light curves of supernovae
Type I.  {\sl Astron. Astrophys. Suppl.} 89: 537--79

\medskip

\noindent Li X--D, van den Heuvel EPJ.  1997.  Evolution of white
dwarf binaries: supersoft X--ray sources and progenitors of Type Ia
		  supernovae.  {\sl Astron. Astrophys.} 322: L9-12

\medskip

\noindent Lira P.  1996.  Light Curves of Supernovae 1990N and 1991T. 
MS thesis, University of Chile

\medskip

\noindent Lira P, Suntzeff NB, Phillips MM, Hamuy M, Maza J, et al.
		  1998.  Optical light curves of the Type Ia
supernovae 1990N and 1991T.  {\sl Astron. J.} 000: 000--

\medskip

\noindent Liu W, Jeffery DJ, Schultz DR.  1997a.  Nebular spectra of
Type Ia supernovae.  {\sl Ap. J. Lett.} 483: L107--10

\medskip

\noindent Liu W, Jeffery DJ, Schultz DR.  1997b.  Nebular spectra of
the unusual Type Ia supernova 1991T.  {\sl Ap. J. Lett.} 486: L35--8

\medskip

\noindent Liu W, Jeffery DJ, Schultz DR, Quinet P, Shaw J, Pindzola
MS.  1997c.  Emission from Cobalt in Type~Ia Supernovae.  {\sl
Ap. J. Lett.} 489: L141--3

\medskip

\noindent Livio M, Donahue M, Panagia N. eds. 1997.  {\sl The
		  Extragalactic Distance Scale}.  Cambridge: Cambridge
		  University.

\medskip

\noindent Livne E. 1990.  Successive detonations in accreting white
dwarfs as an alternative mechanism for Type Ia supernovae.  {\sl
Ap. J.} 354: L53--5

\medskip

\noindent Livne E, Arnett WD.  1995.  Explosions of sub--Chandrasekhar
mass white dwarfs in two dimensions.  {\sl Ap. J.} 452: 62--74

\medskip

\noindent Lundqvist P, Cumming RJ.  1997.  Supernova progenitor
constraints from circumstellar interaction: Type Ia.  In {\sl Advances
		  in Stellar Evolution}, ed. RT Rood, A Renzini,
		  p. 293--7. Cambridge: Cambridge University

\medskip

\noindent Madore BF, Freedman WL. 1991.  The Cepheid distance scale.
{\sl Publ. Astr. Soc. Pac.}  103: 933--57

\medskip

\noindent Madore, BF, Freedman, WL. 1998.  Hipparcos parallaxes and
the Cepheid distance scale.  {\sl Ap. J. Lett.} 000: 000--

\medskip

\noindent Mazzali PA, Danziger IJ, Turatto M. 1995.
A study of the properties of the peculiar SN Ia 1991T through models
of its evolving early--time spectra.  {\sl Astron. Astrophys.} 297:
509--34

\medskip

\noindent Mazzali PA, Chugai N, Turatto M, Lucy LB, Danziger IJ, et
al.  1997. The properties of the peculiar Type Ia supernova 1991bg.
II. The amount of $^{56}$Ni and the total ejecta mass determined from
spectrum synthesis and energetics considerations.  {\sl MNRAS} 284:
151--71

\medskip

\noindent Mazzali PA, Lucy LB, Danziger IJ, Gouiffes C,
		  Cappellaro E, Turatto M.  1993.  Models for the
early--time spectral evolution of the standard type Ia supernova
1990N.  {\sl Astron. Astrophys.} 269: 423--45

\medskip

\noindent Meikle WPS, Cumming RJ, Geballe TR, Lewis JR, Walton
		  NA, et al. 1996.  An early--time infrared and
optical study of the Type Ia supernovae 1994D and 1991T.  {\sl MNRAS}
281: 263--80

\medskip

\noindent Milne PA, The L--S, Leising MD.  1998.  Is positron escape
seen in the light curves of Type Ia supernovae?  {\sl Ap. J.} 000:
000--

\medskip

\noindent Minkowski R. 1939.  The spectra of the supernovae in IC 4182
and NGC 1003.  {\sl Ap. J.} 89: 156--217

\medskip

\noindent Mochkovitch R, Guerrero J, Segretain L.  1997. The merging
of white dwarfs.  See Ruiz--Lapuente, Canal, Isern 1997, pp. 187--204

\medskip

\noindent Mochkovitch R, Livio M. 1990.  The coalescence of white
dwarfs and Type I supernovae.  The merged configuration. {\sl
Astron. Astrophys.} 236: 378--84

\medskip

\noindent Nomoto K, Thielemann F--K, Yokoi K. 1984.  Accreting white
dwarf models for Type Ia supernovae.  III.  Carbon deflagration
supernovae.  {\sl Ap. J.} 286: 644--58

\medskip

\noindent Nomoto K, Iwamoto K, Nakasato N, Thielemann F.--K.,
		  Brachwitz F. et al. 1997.  Type Ia supernovae:
progenitors and constraints on nucleosynthesis.  See Ruiz--Lapuente,
Canal, Isern, 1997, pp. 349--78

\medskip

\noindent Nugent P, Baron E, Branch D, Fisher A, Hauschildt PH.
		  1997.  Synthetic spectra of hydrodynamical models of
Type Ia supernovae.  {\sl Ap. J.} 485: 812--19

\medskip

\noindent Nugent P, Baron E, Hauschildt PH, Branch D.  1995a.
Spectrum synthesis of the Type Ia supernovae 1992A and 1981B. 
		 {\sl  Ap. J.} 441: L33--6

\medskip

\noindent Nugent P, Branch D, Baron E, Fisher A, Vaughan TE,
		  Hauschildt PH. 1995c.  Low Hubble constant from the
physics of Type Ia supernovae.  {\sl Phys. Rev. Lett.} 75: 394--7;
erratum: 75: 1874

\medskip

\noindent Nugent P, Phillips M, Baron E, Branch D,
Hauschildt PH. 1995b.  Evidence for a spectroscopic sequence among
Type Ia supernovae.  {\sl Ap. J. Lett.}  455: L147--50

\medskip

\noindent Patat F, Benetti S, Cappellaro E, Danziger IJ, Della
		  Valle M, et al.  1996.  The Type Ia suppernova 1994D
in NGC 4526: the early phases.  {\sl MNRAS} 278: 111--24

\medskip

\noindent Patat F, Barbon R, Cappellaro E, Turatto M.  1997.  Revised
photometry and color distribution of Type Ia supernovae observed at
Asiago in the seventies.  {\sl Astron. Astrophys.} 317: 423--31

\medskip

\noindent Pauldrach AWA, Duschinger M, Mazzali PA, Puls J, Lennon M,
Miller DL.  1996.  NLTE models for synthetic spectra of Type Ia
supernovae.  {\sl Astron. Astrophys.} 312: 525--538

\medskip

\noindent Perlmutter S, Aldering G, Della Valle M, Deustua S,
		  Ellis RS, et al. 1997c.  Discovery of a Supernova
Explosion at Half the Age of the Universe and its Cosmological
Implications.  {\sl Nature} 000: 000--

\medskip

\noindent Perlmutter S, Deustua S, Goldhaber G, Groom D, Hook I, et
al.  1997a.  Supernovae. {\sl IAU Circ.} No. 6621

\medskip

\noindent Perlmutter S, Gabi S, Goldhaber S, Groom DE, Hook IM,
et al.  1997b.  Measurements of the cosmological parameters $\Omega$
and $\Lambda$ from the first seven supernovae at $z \ge 0.35$.
{\sl Ap. J.} 483: 565--81

\medskip

\noindent Phillips MM.  1993.  The absolute magnitudes of Type Ia
supernovae.  {\sl Ap. J. Lett.} 413: L105--8

\medskip 

\noindent Phillips MM, Phillips AC, Heathcote SR, Blanco VM,
		  Geisler D.  1987.  The Type Ia supernova 1986G in
NGC 5128: optical photometry and spectroscopy.
{\sl Publ. Astr. Soc. Pac.} 99: 592--605

\medskip

\noindent Phillips MM, Wells LA, Suntzeff NB, Hamuy M, Leibundgut
		  B, et al. 1992.  SN 1991T: further evidence of the
heterogeneous nature of Type Ia supernovae.  {\sl Astron. J.} 103: 1632--7

\medskip

\noindent Pierce MJ, Jacoby GH.  1995.  ``New'' $B$ and $V$ photometry
of the ``old'' Type Ia supernovae 1937C: implications for $H_0$.
{\sl Astron. J.} 110: 2885--2909

\medskip

\noindent Pinto PA.  1997.  Time dependence and the opacity of Type Ia
supernovae.  See Ruiz--Lapuente, Canal, Isern 1997, pp. 607--25

\medskip

\noindent Pskovskii YP.  1977.  Brightness, color, and expansion
velocity of SNe I as functions of the rate of brightness decline.
{\sl Sov. Astron. AJ.}  21: 675

\medskip

\noindent Rappaport SA, Di Stefano R, Smith M. 1994.  Formation and
evolution of supersoft X--ray sources.  {\sl Ap. J.}  426: 692--703

\medskip

\noindent Rasio FA, Shapiro SL. 1995.  Hydrodynamics of binary
coalescence.  II.  Polytropes with $\Gamma = 5/3$.  {\sl Ap. J.} 438:
887--903

\medskip

\noindent Renzini A. 1996.  Searching for Type Ia supernova
progenitors.  In {\sl Supernovae and Supernova Remnants}, ed. R
		  McCray, Z. Wang.  Cambridge: Cambridge University,
		  pp. 77--85

\medskip

\noindent Richmond MW, Treffers RR, Filippenko AV, Van Dyk SD,
		  Paik Y, et al. 1995.  $UBVRI$ photometry of the Type
Ia SN 1994D in NGC 4526.  {\sl Astron. J.} 109: 2121--33
		
\medskip

\noindent Riess, AG. 1996. Type Ia supernova multicolor light curve
shapes.  PhD thesis. Harvard University
  
\medskip

\noindent Riess AG, Davis M, Baker J, Kirshner RP. 1997b.  The
velocity field from Type Ia supernovae matches the gravity field from
galaxy surveys.  {\sl Ap. J.} 488: L1--5.

\medskip

\noindent Riess AG, Filippenko AV, Leonard DC, Schmidt BP, Suntzeff
		  N, et al. 1997a. Time dilation from spectral feature
age measurements of Type Ia supernovae.  {\sl Astron. J.} 114: 722--9

\medskip

\noindent Riess AG, Nugent P, Filippenko AV, Kirshner RP, Perlmutter
S.  1998.  Snapshot distances to SNe~Ia --- all in ``one'' night's
work.  ApJ: 000: 000--

\medskip

\noindent Riess AG, Press WH, Kirshner RP.  1995a.  Using Type Ia
supernova light curve shapes to measure the Hubble constant.
{\sl Ap. J. Lett.} 438: L17--20

\medskip

\noindent Riess AG, Press WH, Kirshner RP.  1995b.  Determining the
motion of the local group using Type Ia supernova light curve shapes.
{\sl Ap. J. Lett.} 445: L91--4

\medskip

\noindent Riess AG, Press WH, Kirshner RP.  1996a.  A precise distance
indicator: Type Ia supernova multicolor light--curve shapes.  {\sl
Ap. J.}  473: 88--109

\medskip

\noindent Riess AG, Press WH, Kirshner RP.  1996b.  Is the dust
obscuring supernovae in distant galaxies the same as the dust in the
Milky Way?  {\sl Ap. J.} 473: 588--94

\medskip

\noindent Ruiz--Lapuente P.  1996. The Hubble constant from
$^{56}$Co--powered nebular candles.  {\sl Ap. J.} 465: L83--6

\medskip

\noindent Ruiz--Lapuente P, Canal R, Burkert A.  1997.  SNe Ia: On the
binary progenitors and expected statistics.  See Ruiz--Lapuente,
		  Canal, Isern 1997, pp. 205--230

\medskip

\noindent Ruiz--Lapuente P, Canal R, Isern J, eds. 1997.  {\sl
Thermonuclear Supernovae}.  Dordrecht: Kluwer

\medskip

\noindent Ruiz--Lapuente P, Cappellaro E, Turatto M, Gouiffes C,
		  Danziger IJ, Della Valle M, Lucy LB. 1992.  Modeling
the iron--dominated spectra of the Type Ia supernova 1991T at
premaximum.  {\sl Ap. J. Lett.} 387: L33-6

\medskip

\noindent Saha A, Labhardt L, Schwengeler H, Macchetto FD,
		  Panagia N, et al.  1994.  Discovery of Cepheids in
IC 4182: absolute peak brightness of SN Ia 1937C and the value of
$H_0$.  {\sl Ap. J.} 425: 14--34

\medskip

\noindent Saha A, Sandage A, Labhardt L, Schwengeler H, Tammann
		  GA, et al.  1995.  Discovery of Cepheids in NGC
5253: absolute peak brightness of SNe Ia 1895B and 1972E and the value
of $H_0$.  {\sl Ap. J.}  438: 8--26

\medskip

\noindent Saha A, Sandage A, Labhardt L, Tammann GA, Macchetto
		  FD, Panagia N.  1996a.  Cepheid calibration of the
peak brightness of SNe Ia.  V.  SN 1981B in NGC 4536.  {\sl Ap. J.} 466:
55--91

\medskip

\noindent Saha A, Sandage A, Labhardt L, Tammann GA, Macchetto
		  FD, Panagia N. 1996b.  Cepheid calibration of the
peak brightness of Type Ia supernovae.  VI. SN 1960F in NGC 4496A.
{\sl Ap. J. Suppl.} 107: 693

\medskip

\noindent Saha, A, Sandage, A, Labhardt, L, Tammann, GA, Macchetto,
		  FD, Panagia, N.  1997.  Cepheid calibration of the
peak brightness of Type Ia supernovae.  VII.  SN 1990N in NGC 4639.
{\sl Ap. J.} 486: 1--20

\medskip

\noindent Saio H, Gautschy A. 1998.  The Cepheid Metallicity
Dependence.  {\sl Ap. J} 000: 000--

\medskip

\noindent Sandage A. 1997. in {\sl Annual Report of the Carnegie
Observatories}. {\sl B.A.A.S.} 28: 52--2

\medskip

\noindent Sandage A, Saha A, Tammann GA, Panagia N, Macchetto
		  FD. 1992.  The Cepheid distance to IC 4182;
calibration of $M_V$ for SN Ia 1937C and the value of $H_0$.
{\sl Ap. J. Lett.} 401: L7--10

\medskip

\noindent Sandage A, Tammann GA. 1982.  Steps toward the Hubble
constant.  VIII.  The global value.  {\sl Ap. J.} 256: 339--45

\medskip

\noindent Sandage A, Tammann GA. 1993.  The Hubble diagram in $V$ for
supernovae of Type Ia and the value of $H_0$ therefrom.  {\sl Ap. J.}
415: 1--9

\medskip

\noindent Sandage A, Tammann GA. 1997.  The evidence for the long
distance scale with $H_0 < 65$.  In {\sl Critical Dialogues in
Cosmology}, ed. Turok N, pp. 130--55.  Singapore: World Scientific

\medskip

\noindent Sandage A, Tammann GA. 1998.  Confirmation of previous
ground--based Cepheid P--L zero--points using Hipparcos trigonometric
parallaxes.  {\sl MNRAS} 000: 000--

\medskip

\noindent Sasselov DD, Beaulieu JP, Renault C, Grison P, Ferlet
		  R, et al.  1997.  Metallicity effects on the Cepheid
extragalactic distance scale from EROS photometry in the Large
Magellanic Cloud and the Small Magellanic Cloud.  {\sl Astron. Ap.}
324: 471--82

\medskip

\noindent Schaefer BE. 1994.  The peak brightness of SN 1937C in IC
4182 and the Hubble constant.  {\sl Ap. J.} 426: 493--501

\medskip

\noindent Schaefer BE. 1995a.  The peak brightness of SN 1895B in NGC
5253 and the Hubble constant.  {\sl Ap. J. Lett.} 447: L13--6

\medskip

\noindent Schaefer BE. 1995b.  The peak brightness of supernovae in
the $U$ band and the Hubble constant.  {\sl Ap. J. Lett.} 450: L5--9

\medskip

\noindent Schaefer BE. 1995c.  The peak brightness of SN 1981B in NGC
4536 and the Hubble constant.  {\sl Ap. J. Lett.} 449: L9--12

\medskip

\noindent Schaefer BE. 1996a.  The peak brightness of SN 1937C and the
Hubble constant: comments on Pierce \& Jacoby (1995).  {\sl
Astron. J.} 111: 1668--1674

\medskip

\noindent Schaefer BE. 1996b.  The peak brightness of SN 1960F in NGC
4496 and the Hubble constant.  {\sl Ap. J. Lett.} 460: L19--23

\medskip

\noindent Schlegel EM, Petre R. 1993.  A {\sl ROSAT} upper limit on
X--ray emission from SN~1992A.  {\sl Ap. J. Lett.} 412: L29--32

\medskip

\noindent Schmidt B, Phillips M, Hamuy M, Avil\'es R, Suntzeff N, et
al.  1997.  Supernovae. {\sl IAU Circ.} No. 6646

\medskip

\noindent Shigeyama T, Nomoto K, Yamaoka H, Thielemann F--K.  1992.
Possible Models for the Type~Ia Supernova 1990N.  {\sl Ap. J. Lett.}
386: L13--6

\medskip

\noindent Stothers RB.  1988.  Abundance Effects on the Cepheid
Distance Scale.  {\sl Ap. J.} 329: 712--9

\medskip

\noindent Tammann GA, Federspiel M.  1997.  Focusing in on $H_0$.  See
Livio, Donahue, Panagia 1997, pp. 137--70

\medskip

\noindent Tammann GA, Sandage A.  1995.  The Hubble diagram for
supernovae of Type Ia.  II.  The effects on the Hubble constant of a
correlation between absolute magnitude and light curve decay rate.
{\sl Ap. J.} 452: 16--24

\medskip

\noindent Teerikorpi P. 1997.  Observational selection bias affecting
the determination of the extragalactic distance scale.  {\sl
Annu. Rev. Astron. Astrophys.} 35: 101--36

\medskip

\noindent Tripp R. 1997.  Using distant Type Ia supernovae to measure
the cosmological expansion parameters.  {\sl Astron. Astrophys.} 
325: 871--6

\medskip

\noindent Tripp R. 1998.  A two--parameter luminosity correction for
Type Ia supernovae.  {\sl Astron. Astrophys.} 000: 000--

\medskip

\noindent Turatto M., Benetti S, Cappellaro E, Danziger IJ, Della
		  Valle M, et al. 1996.  The properties of the
peculiar Type Ia supernova 1991bg.  I.  Analysis and discussion of two
years of observations.  {\sl MNRAS} 283: 1--17

\medskip

\noindent Tutukov A, Yungelson LR. 1994.  Merging of binary white
dwarfs, neutron stars, and black holes under the influence of
gravitational wave radiation.  {\sl MNRAS} 268: 871--9

\medskip

\noindent Vacca WD, Leibundgut B.  1996.  The rise times and
bolometric light curve of SN 1994D: constraints on models of Type Ia
supernovae.  {\sl Ap. J. Lett.} 471, L37--40

\medskip

\noindent van den Bergh S. 1995.  A new method for the determination
of the Hubble parameter.  {\sl Ap. J. Lett.} 453: L55--56

\medskip

\noindent van den Bergh S. 1996.  The luminosities of supernovae of
Type Ia derived from Cepheid calibrations.  {\sl Ap. J.} 472: 431--4

\medskip

\noindent van den Heuvel EPJ, Bhattacharya D, Nomoto K, Rappaport SA.
1992.  Accreting White Dwarf Models for CAL83, CLA87, and Other
Ultrasoft X--ray Sources in the LMC.  {\sl Astron. Astrophys.} 262:
97--105

\medskip

\noindent Vaughan TE, Branch D, Miller DL, Perlmutter S.  1995.
The blue and visual absolute magnitude distributions of Type Ia
supernovae.  {\sl Ap. J.} 439: 558--64

\medskip

\noindent von Hippel T, Bothun GD, Schommer RA. 1997.  Stellar
populations and the white dwarf mass function: connections to
supernova Ia luminosities.  {\sl Astron. J.} 114: 1154--64

\medskip

\noindent Wang L, H\"oflich P, Wheeler JC.  1997.  Supernovae and
their host galaxies.  {\sl Ap. J.} 483: L29--32.

\medskip

\noindent Wang L, Wheeler JC, H\"oflich P.  1997.  Polarimetry of the
Type Ia supernova 1996X.  {\sl Ap. J.} 476: L27--30

\medskip

\noindent Wang L, Wheeler JC, Li Z, Clocchiatti A.  1996.  Broadband
polarimetry of supernovae: 1994D, 1994Y, 1994ae, 1995D, and 1995H.
{\sl Ap. J.} 467: 435--45

\medskip

\noindent Wells LA, Phillips MM, Suntzeff NB, Heathcote SR, Hamuy M,
		  et al. 1994.  The Type Ia supernova 1989B in NGC
3627 (M66).  {\sl Astron. J.} 108: 2233--50

\medskip

\noindent Wheeler JC, Harkness RP, Khokhlov AM, H\"oflich P.  1995.
Stirling's supernovae: a survey of the field.  {\sl Phys. Reports}
		  256: 211--35.

\medskip

\noindent Woosley SE, Weaver TA. 1986.  The physics of supernova 
explosions.  {\sl Annu. Rev. Astron. Astrophys.} 24: 205--53

\medskip

\noindent Woosley SE, Weaver TA.  1994a.  Sub--Chandrasekhar models
for Type Ia supernovae.  {\sl Ap. J.} 423: 371--9

\medskip

\noindent Woosley, SE, Weaver TA.  1994b.  Massive stars, supernovae,
and nucleosynthesis.  In {\sl Supernovae}, ed. S Bludman, R
		  Mochkovitch, J Zinn--Justin.  Elsevier Science,
		  Amsterdam, pp. 63--154

\medskip

\noindent Yungelson LR, Livio M. 1998.  Type~Ia Supernovae: An
Examination of Potential Progenitors and the Redshift Distribution.
{\sl Ap. J.} 000: 000--

\medskip

\noindent Yungelson LR, Livio M, Truran JW, Tutukov A, Fedorova A.
		  1996.  A model for the Galactic population of binary
supersoft X--ray sources.  {\sl Ap. J.} 466: 890--910

\medskip

\noindent Zehavi I, Riess AG, Kirshner RP, Dekel A.  1998.  A local
Hubble bubble from SNe~Ia?  {\sl Ap. J.} 000: 000--

\vfill\eject\end